\documentclass[letterpaper,onecolumn,allowtoday,accepted=2023-03-26,preprintnumbers,10pt]{quantumarticle}
\pdfoutput=1
\usepackage[numbers,sort&compress]{natbib}
\usepackage[utf8]{inputenc}
\usepackage[english]{babel}
\usepackage[T1]{fontenc}
\usepackage{textgreek}
\usepackage{braket}
\usepackage{graphicx,subfigure}
\usepackage{physics}
\usepackage{multirow}
\usepackage{siunitx}

\usepackage{tikz}
\usetikzlibrary{decorations.markings, arrows, positioning, calc}

\usepackage{graphicx,times}
\usepackage{latexsym}
\usepackage{mathtools}

\usepackage{amsmath,amssymb,amsbsy,amsfonts}
\usepackage{array}
\usepackage{bm}

\usepackage{graphics}
\graphicspath{{./fig/}}

\usepackage{mathrsfs}
\usepackage{xcolor}
\usepackage{cancel}
\usepackage[normalem]{ulem}

\usepackage[unicode, pdfprintscaling=None, colorlinks]{hyperref}
\hypersetup{
    colorlinks,
    allcolors=blue!50!black,
}

\usepackage[capitalise]{cleveref}

\usepackage{relsize}

\usepackage{microtype}

\usepackage{xfrac}
\usepackage{orcidlink}

\usepackage{ulem} 
\usepackage{amssymb}

\newcommand\del\partial

\usepackage{graphicx}% Include figure files
\usepackage{dcolumn}% Align table columns on decimal point
\usepackage{bm}% bold math

\begin{document}

\title{State Preparation in the Heisenberg Model through Adiabatic Spiraling}

 \author{Anthony N. Ciavarella \,\orcidlink{0000-0003-3918-4110}}
 \email{aciavare@uw.edu}
 \affiliation{InQubator for Quantum Simulation (IQuS), Department of Physics, University of Washington, Seattle, Washington 98195-1550, USA}
 \author{Stephan Caspar\,\orcidlink{0000-0002-3658-9158}}
 \email{caspar@uw.edu}
 \affiliation{InQubator for Quantum Simulation (IQuS), Department of Physics, University of Washington, Seattle, Washington 98195-1550, USA}
 \author{Marc Illa\,\orcidlink{0000-0003-3570-2849}}
 \email{marcilla@uw.edu}
 \affiliation{InQubator for Quantum Simulation (IQuS), Department of Physics, University of Washington, Seattle, Washington 98195-1550, USA}
 \author{Martin J. Savage\,\orcidlink{0000-0001-6502-7106}}
 \email{mjs5@uw.edu}
 \affiliation{InQubator for Quantum Simulation (IQuS), Department of Physics, University of Washington, Seattle, Washington 98195-1550, USA}
%\preprint{IQuS@UW-21-033}
\date{\today}

\begin{abstract}
An adiabatic state preparation technique, called the adiabatic spiral, is proposed for the Heisenberg model. This technique is suitable for implementation on a number of quantum simulation platforms such as Rydberg atoms, trapped ions, or superconducting qubits. Classical simulations of small systems suggest that it can be successfully implemented in the near future. A comparison to Trotterized time evolution is performed and it is shown that the adiabatic spiral is able to outperform Trotterized adiabatics.
\end{abstract}

\maketitle

%\tableofcontents

\section{Introduction}
\noindent
Quantum simulations of lattice gauge theories, such as quantum chromodynamics (QCD) and quantum electrodynamics (QED),  
are anticipated, in the future,  
to enable reliable predictions for non-equilibrium dynamical processes, ranging from 
fragmentation in high-energy hadronic collisions, 
through to transport in extreme astrophysical environments.
While a quantum advantage has yet to be established for a scientific application, 
including quantum field theories, 
there are substantial efforts underway to perform quantum simulations that can be compared with experiment, or impact future experiments,
and impressive progress has been made toward these objectives in the last decade. This includes the development of techniques to simulate abelian gauge theories~\cite{Zohar:2011cw,Zohar:2012ay,Tagliacozzo:2012vg,Zohar:2012ts,PhysRevLett.110.125303,Hauke:2013jga,Marcos:2014lda,Kuno:2014npa,Bazavov:2015kka,Kasper:2015cca,Brennen:2015pgn,Kuno:2016xbf,Martinez:2016yna,Kasper:2016mzj,Muschik:2016tws,Gonzalez-Cuadra:2017lvz,Klco:2018kyo,Kaplan:2018vnj,Kokail:2018eiw,Stryker:2018efp,Davoudi:2019bhy,Avkhadiev:2019niu,Magnifico:2019kyj,Luo:2019vmi,Kharzeev:2020kgc,Shaw2020quantumalgorithms,PhysRevLett.122.050403,Yang_2020,Bender:2020jgr,Haase:2020kaj,Halimeh:2020ecg,Robaina:2020aqh,Paulson:2020zjd,VanDamme:2020rur,Ott:2020ycj,Bauer:2021gup,Kan:2021nyu,Aidelsburger:2021mia,Meurice:2021pvj,Mueller:2021gxd,Riechert:2021ink,Halimeh:2021lnv,Zhang:2021bjq,Gustafson:2021jtq,Thompson:2021eze,Kan:2021blb,Ashkenazi:2021ieg,Bauer:2021gek,Nguyen:2021hyk,Halimeh:2022rwu,Mildenberger:2022jqr,Halimeh:2022pkw,Greenberg:2022kzy,Grabowska:2022uos,Davoudi:2022uzo,https://doi.org/10.48550/arxiv.2206.12454}, non-abelian gauge theories~\cite{Brower:1997ha,PhysRevA.73.022328,Banerjee:2012xg,Tagliacozzo:2012df,Zohar:2012xf,Wiese:2013uua,Zohar:2016iic,Banuls:2017ena,Alexandru:2019nsa,Klco:2019evd,Banuls:2019bmf,Ji:2020kjk,Halimeh:2020djb,Kasper:2020owz,Ciavarella:2021nmj,ARahman:2021ktn,Atas:2021ext,Davoudi:2021ney,Stryker:2021asy,Zohar:2021nyc,Halimeh:2021vzf,Wiese:2021djl,Alam:2021uuq,Funcke:2021aps,VanDamme:2021njp,Alexandrou:2021ynh,Ciavarella:2021lel,Hartung:2022hoz,Illa:2022jqb,Ji:2022qvr,Carena:2022kpg,Ciavarella:2022zhe,Bauer:2022hpo,Raychowdhury:2022wbi,Rahman:2022rlg,Farrell:2022wyt,Atas:2022dqm,Carena:2022hpz,Gustafson:2022xdt,Avkhadiev:2022ttx,Farrell:2022vyh}, fermionic field theories~\cite{Mishra:2019xbh,Perlin:2021xux,Bringewatt:2022zgq,Asaduzzaman:2022bpi}, and scalar field theories~\cite{Yeter-Aydeniz:2018mix,Klco:2019xro,Klco:2019yrb,Barata:2020jtq,Yeter-Aydeniz:2021mol,Deliyannis:2021che,Caspar:2022llo}. There has also been the development of techniques to extract observables of interest to nuclear physics~\cite{Dumitrescu:2018njn,Lu:2018pjk,Cloet:2019wre,Shehab:2019gfn,Mueller:2020vha,Knaute:2021xna,Yeter-Aydeniz:2021olz}, scattering processes in high energy physics~\cite{Gustafson:2019mpk,Bauer:2019qxa,Gustafson:2021mky,Yeter-Aydeniz:2020jte,Milsted:2020jmf,Gustafson:2021imb,Deliyannis:2022uyh,Dreher:2022scr} and methods to mitigate errors on noisy quantum hardware~\cite{Wang:2021iox,Iannelli:2021jhs,Yeter-Aydeniz:2022vuy,Halimeh:2022mct,Tuysuz:2022knj,Jang:2022nun}.
Currently available hardware and our present understanding of quantum algorithms has 
so far limited quantum simulations of lattice gauge theories to one- and two-dimensions 
with only a small number of lattice sites~\cite{Klco:2019evd,Atas:2021ext,Ciavarella:2021nmj,Ciavarella:2021lel,Illa:2022jqb,Rahman:2022rlg,Farrell:2022wyt,Atas:2022dqm}.
Qualitative insights can be gained from simulations of spin models that share one or more features of QCD, QED or low-energy effective field theories (EFTs) relevant to nuclear and particle physics.  
These include models that are in the same universality class as these theories, that can be fruitfully digitized onto qubit registers or mapped to analog quantum simulators, 
such as arrays of Rydberg atoms.

The Heisenberg model with arbitrary couplings is computationally universal in the sense that all other lattice models can be simulated in arbitrary dimensions, particle content and interactions by simulations of Heisenberg models~\cite{Cubitt_2018}.
Therefore, detailed understandings of quantum simulations of the Heisenberg model inform the simulations of quantum field theories describing the forces of nature.
Translating results obtained from lattice field theories to predictions that 
can be compared with experiment requires that all relevant physical length scales are much larger than the scale of discretization of spacetime, and 
universality guarantees that low-energy continuum physics 
can be reproduced from simulations that are tuned near a 2nd order critical point~\cite{WILSON197475,RevModPhys.55.583,RevModPhys.51.659}. 
As an example, it has been proposed that 
universality assures that 
the continuum physics of the $1+1$d O(3) NL$\sigma$M,
which has been studied as a toy model of quantum chromodynamics (QCD) 
due to sharing a number of qualitative features such as asymptotic freedom, dynamical transmutation, the generation of a
non-perturbative mass gap and non-trivial $\theta$ vacua, 
can be recovered from simulations of an anti-ferromagnetic Heisenberg model~\cite{CHANDRASEKHARAN2002388,BROWER2004149,PhysRevD.99.074501,PhysRevD.100.054505,PhysRevLett.126.172001,https://doi.org/10.48550/arxiv.1911.12353,https://doi.org/10.48550/arxiv.2203.00059,https://doi.org/10.48550/arxiv.2203.15766}. 
Thus, quantum simulations of the low-energy dynamics of the anti-ferromagnetic Heisenberg model, which requires preparing a low-energy state and evolving it forward in time, 
are expected to provide key insights into strategies for simulating QCD, including state preparation.

To enable practical quantum simulations of physical systems, preparation of states that have energies much less than the inverse lattice spacing is required. One proposal for preparing low energy states in both digital and analog quantum simulation is adiabatic switching. This works by beginning in the ground state of a known Hamiltonian and slowly varying the Hamiltonian through a path in parameter space where the energy gap does not close. Implementation of adiabatic switching in a quantum simulation requires the ability to prepare the eigenstate of the initial Hamiltonian and simulate time evolution. Schemes for simulating the Heisenberg model's time evolution have been proposed using digital quantum simulation~\cite{Childs_2018,Childs_2021}, hybrid digital-analog simulation~\cite{PRXQuantum.2.020328}, periodically driven trapped ions~\cite{PhysRevB.95.024431}, global microwave pulses on Rydberg atoms~\cite{PRXQuantum.3.020303}, nuclear spins~\cite{MADI1997300,PhysRevLett.124.030601}, and adding strong single qubit terms to systems described by Ising interactions~\cite{https://doi.org/10.48550/arxiv.2207.09438}. 

In this work, we present an argument that the eigenstates of the Heisenberg model are approximate eigenstates of the Ising model with a strong external field pointed in an appropriate direction. This is used to develop an analogue quantum simulation technique, called the adiabatic spiral, to adiabatically prepare the ground state of the Heisenberg model by adiabatically varying the direction of the external field in the Ising model. The feasibility of implementing the adiabatic spiral on Rydberg atoms and D-Wave's quantum annealer is investigated. It is found that current Rydberg atom experiments have sufficient coherence time and drive field strengths to implement the adiabatic spiral, while the D-Wave quantum annealer suffers from some limitations.

%%%%%%%%%%%%%%%%%%%%%%%%%%%%
\section{Adiabatic Spirals: Spiraling Toward Ground States}
\begin{figure}[!ht]
    \centering

    \subfigure[Laboratory]{\label{fig:LabSpiral}\includegraphics[width=6cm,trim=0.8cm 2.3cm 0.8cm 1.cm]{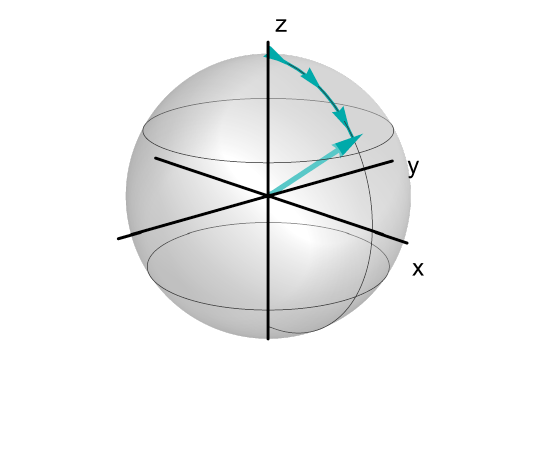}}
    \subfigure[Interaction Picture]{\label{fig:IntSpiral}\includegraphics[width=6cm,trim=0.8cm 2.3cm 0.8cm 1.cm]{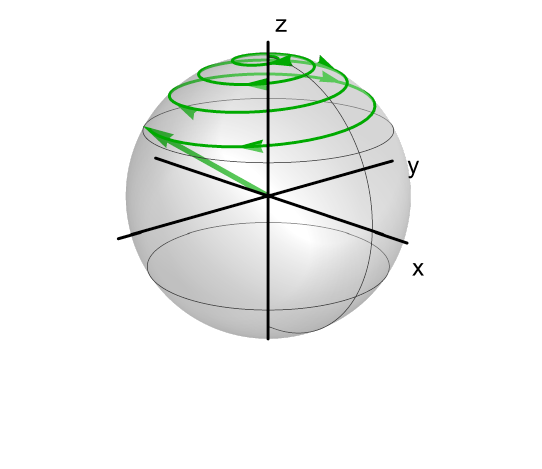}}

    \caption{Paths on the unit sphere taken by (a) $\Vec{h}(t)$ (the time-dependent drive field defined in Eq.~(\ref{eq:FreeHam}))
    and (b) $\Vec{e}(t)$ (defining the time dependence of the $\hat{Z}$ operator in the interaction picture) during the course of the adiabatic spiral. $\theta(t)$ is taken to be $\theta(t)=\arccos\left(\frac{1}{\sqrt{3}}\right)\frac{t}{T}$ and $t$ varies from $t=0$ to $t=T$. During the evolution, $\Vec{e}(t)$ is precessing around $\Vec{h}(t)$ while the opening angle $\theta$ changes adiabatically, resulting in a spiral path on the unit sphere in the interaction picture. Vectors from the origin indicate the direction at the end of the spiral evolution.}
    \label{fig:Zpath}
\end{figure}
\noindent
We recently showed that the Ising model with large external fields in the transverse and longitudinal directions can approximate the time evolution of the Heisenberg model at discrete time intervals~\cite{https://doi.org/10.48550/arxiv.2207.09438}. 
This generalizes previous work showing the Ising model with a large transverse field approximates the dynamics of the 
XY model~\cite{Richerme_2014,Jurcevic_2014,Wall_2017,Kiely_2018,https://doi.org/10.48550/arxiv.2208.01869}, 
and is related to techniques used to study pre-thermalization~\cite{abanin2017rigorous,PhysRevX.7.011026,PhysRevB.100.020301,https://doi.org/10.48550/arxiv.2103.07485}.
Explicitly, 
if a quantum simulator evolves under the Hamiltonian
\begin{equation} 
\label{eq:IsingResource}
\hat{H}^{\text{Ising}} = \sum_{i,j} J_{i,j} \hat{Z}_i \hat{Z}_{j} + \sum_i \frac{\Omega}{2} \left( \cos{\theta}\hat{Z}_i + \sin{\theta} \hat{X}_i \right) 
\ \ \ ,
\end{equation}
then the time evolution of 
\begin{equation}
    \hat{H}^{\text{Heis.}} = \sum_{i,j} J_{i,j} \left[ \cos^2{\theta} \hat{Z}_i \hat{Z}_j +\frac{\sin^2{\theta}}{2}\left(\hat{X}_i \hat{X}_j + \hat{Y}_i \hat{Y}_j\right) \right]
    \label{eq:GenericHeis}
\end{equation}
will be approximated at times that are integer multiples of $t=\frac{2 \pi}{\Omega}$, 
up to a change of basis and corrections that are $O\left(\frac{1}{\Omega}\right)$. In this work, $\hat{X}$, $\hat{Y}$ and $\hat{Z}$ refer to the respective Pauli operators.

We now observe that if the time evolution was reproduced exactly at these time intervals, it would guarantee that $\hat{H}^{\text{Heis.}}$ and $\hat{H}^{\text{Ising}}$ share the same eigenstates and that their energy levels agree up to integer multiples of $\Omega$. This would suggest that by beginning with $\theta=0$ and adiabatically increasing $\theta$, it should be possible to prepare an eigenstate of the Heisenberg model from an eigenstate of the Pauli $\hat{Z}$ operators. When viewed in the interaction picture where the free part of the Hamiltonian is taken to be the single spin driving terms
\begin{equation}
H_0(t) = \frac{\Omega}{2} \sum_i \vec{h}(t) \cdot \vec{\sigma}_i \ \ \ ,  
\label{eq:FreeHam}
\end{equation}
the local $\hat{Z}$ operator becomes
\begin{equation}
    \hat{Z}^I_j(t) = U_0^\dagger(t) \hat{Z}_j U_0(t) = \Vec{e}(t) \cdot \Vec{\sigma}_j
    \ \ ,
    \label{eq:Zinteraction}
\end{equation}
where $\Vec{\sigma}_j$ is a vector of Pauli matrices and $\Vec{e}(t)$ is a unit vector. For the Hamiltonian in Eq.~(\ref{eq:IsingResource}), $\Vec{e}(t)$ rotates in a circle about a vector pointing in the $\cos \theta \hat{z} + \sin\theta \hat{x}$ direction. As $\theta$ is adiabatically varied to prepare the ground state of a Heisenberg model, $\Vec{e}(t)$ will move in a spiral motion along the surface of the sphere as shown in Fig. \ref{fig:Zpath}. For this reason we call this method of state preparation adiabatic spiraling. It is important to note that the implementation of the adiabatic spiral does not require the switching time to be an integer multiple of $\frac{2\pi}{\Omega}$. Studying the evolution of the Ising model at periodic times was only necessary to argue that the eigenstates of the Heisenberg model are also eigenstates of the Ising model up to $O\left(\frac{1}{\Omega}\right)$ corrections. Typically in quantum simulation, adiabatic switching is done between ground states of gapped Hamiltonians. In contrast, the adiabatic spiral can be understood as performing adiabatic switching in the middle of the spectrum of the Ising model to prepare eigenstates of the Heisenberg model.

In practice, the initial eigenstate of the Ising model will often be degenerate, and this degeneracy will need to be split for the adiabatic approximation to be valid. This can be done by modifying the single-spin terms in the Hamiltonian. Explicitly, at times that are integer multiples of $t=\frac{2 \pi}{\Omega}$, the time evolution generated by
\begin{equation}
    \hat{H}^{\text{Ising}} = \sum_{i,j} J_{i,j} \hat{Z}_i \hat{Z}_{j} + \sum_j \frac{\Omega}{2} \left( \cos{\theta}\hat{Z}_j + \sin{\theta} \hat{X}_j \right) + \frac{h_P(j)}{2} \hat{Z_j}
    \ \ ,
\end{equation}
can be approximated by the Floquet operator,
\begin{equation}
\begin{split}
\hat{U}_F  =  \hat{U}^\dagger_B\exp \Bigg\{
&-i\frac{2\pi}{\Omega}\sum_{i j} J_{i j} \left(\cos^2{\theta} \hat{Z}_i \hat{Z}_j +\frac{\sin^2{\theta}}{2}\left(\hat{X}_i \hat{X}_j + \hat{Y}_i \hat{Y}_j\right) \right)\\
 & -i\frac{2\pi}{\Omega} 
 \sum_j \frac{1}{2} \cos\theta\  h_P(j)\  \hat{Z}_j 
   + O\left(\frac{1}{\Omega^2}\right) \Bigg\} \hat{U}_B \ \ \ , 
\end{split}
\end{equation}
where $\hat{U}_B$ is a local change of basis given by $\hat{U}_B = \prod_j e^{i \frac{\theta}{2} \hat{Y}_j} $. The additional single-qubit term can be tuned to create an energy penalty that breaks the degeneracy of the initial Hamiltonian which enables the application of the adiabatic approximation. With these additional single-qubit terms present, the same arguments at large $\Omega$ can be used to justify the applicability of the adiabatic spiral.

As an explicit demonstration, we consider the preparation of the ground state of the anti-ferromagnetic Heisenberg model on a 1D chain, with a Hamiltonian of the form
\begin{equation}
    \hat{H}^{\text{Heis.}} = J \sum_j \left[ \hat{X}_j \hat{X}_{j+1} + \hat{Y}_j \hat{Y}_{j+1} + \hat{Z}_j \hat{Z}_{j+1} \right] 
    \label{eq:antiH}
    \ \ \ .
\end{equation}
Preparing the ground state of this system with the adiabatic spiral will require beginning in an eigenstate of the Ising model that is adiabatically connected to the ground state of the Heisenberg model. Equation~\ref{eq:GenericHeis} reproduces Eq.~\ref{eq:antiH} when $\theta=\arccos{\frac{1}{\sqrt{3}}}$ and would suggest that the ground state of Eq.~\ref{eq:antiH} is connected to the ground state of Eq.~\ref{eq:GenericHeis} at other values of $\theta$. The ground state of the Heisenberg Hamiltonian with $\theta=0$ is a state with spins alternating up and down
in the $\hat{z}$ direction (a N\'{e}el state), e.g., $\ket{\uparrow\downarrow\uparrow\downarrow\uparrow\downarrow ...}$. This ground state is degenerate and the degeneracy can be split by adding a single qubit term to the Hamiltonian with alternating signs. Therefore, the ground state of the full Heisenberg model can be prepared by beginning in a N\'{e}el state and applying a time-dependent Hamiltonian of the form
\begin{equation}
     \hat{H}(t) = \sum_j \left[ J \hat{Z}_j \hat{Z}_{j+1} + \frac{\Omega}{2} \left( \frac{1}{\sqrt{3}} \hat{Z}_j + f(t) \hat{X}_j \right) + \frac{h_P(t)}{2} (-1)^j \hat{Z}_j \right] \ \ \ , 
    \label{eq:SpiralHam}
\end{equation}
for a time $T$, where $h_P(0)>0$, $h_P(T)=0$, $f(0)=0$, and $f(T)=\sqrt{\frac{2}{3}}$.
Explicitly, 
if $\ket{Vac}$ is the ground state of the anti-ferromagnetic Heisenberg model given in Eq.~(\ref{eq:antiH}) and $\ket{N\acute{e}el}$ is a N\'{e}el state, then
\begin{equation}
	\ket{Vac} = \mathcal{T} e^{-i \int_0^T dt \hat{H}(t)} \ket{N\acute{e}el}
	\ \ ,
	\label{eq:UnitarySpiral}
\end{equation}
up to $O\left(\frac{1}{\Omega}\right)$ and finite time corrections where $\hat{H}(t)$ is the time-dependent Hamiltonian defined in Eq.~(\ref{eq:SpiralHam}).

Typically, analog quantum simulators 
are initialized with all qubits in their ground state, 
e.g., $|\downarrow\rangle^{\otimes n}$.
However, to apply the adiabatic spiral to the anti-ferromagnetic Heisenberg model, 
relevant for simulations of the O(3) NL$\sigma$M, 
the initial state should be a N\'{e}el state,
which can be accomplished by applying a rotation on every other qubit.
Alternately, computations could be performed in a different basis.
If $\mathbf{X}$ is defined to be a product of Pauli X's on every other site such that $\ket{N\acute{e}el}=\mathbf{X}\ket{\downarrow}^{\otimes n}$ where $\ket{\downarrow}^{\otimes n}$ is the state with all spins down, then Eq.~(\ref{eq:UnitarySpiral}) can be written as
\begin{equation}
	\ket{Vac} = \mathcal{T} e^{-i \int_0^T dt \hat{H}(t)} \mathbf{X}\ket{\downarrow}^{\otimes n} 
	\ \ \ .
\end{equation}
Multiplying both sides of this equation by $\mathbf{X}$ yields
\begin{equation}
	\mathbf{X} \ket{Vac} = \mathbf{X} \mathcal{T} e^{-i \int_0^T dt \hat{H}(t)} \mathbf{X}\ket{\downarrow}^{\otimes n} = \mathcal{T} e^{-i \int_0^T dt \tilde{H}(t)} \ket{\downarrow}^{\otimes n}
	\ \ \ ,
	\label{eq:Xvac}
\end{equation}
where
\begin{equation}
    \tilde{H}(t) = \sum_j \left[ -J\hat{Z}_j \hat{Z}_{j+1} + \frac{\Omega}{2} \left( \frac{(-1)^j}{\sqrt{3}} \hat{Z}_j + f(t) \hat{X}_j \right) + \frac{h_P(t)}{2} \hat{Z}_j \right]
    \label{eq:spiraltilde}
    \ \ \ .
\end{equation}
If the quantum simulator can directly implement $\tilde{H}(t)$, 
then an adiabatic spiral can be used to adiabatically prepare the Heisenberg ground state from the $\ket{\downarrow}^{\otimes n}$ state up to a basis transformation. However, on some analog quantum simulators such as Rydberg atom systems, the sign of the two-spin interaction is fixed to be positive. 
This does not present an issue in using an adiabatic spiral, as it can be implemented with $-\tilde{H}(t)$. 
This is because a change of overall sign does not change the eigenstates or presence of an energy gap between eigenstates, 
indicating that the adiabatic approximation remains valid.

%%%%%%%%%%%%%%%%%%%%%%%%%%%%%%%%%%%%%%%%%%%%%%%%%%%%%%%%%%%%%%%%%%%%%%%%%%%%%%%%%%%%%%%%%%
\section{A Numerical Example}
\begin{figure}[!ht]
    \centering
    \includegraphics[width=8.6cm]{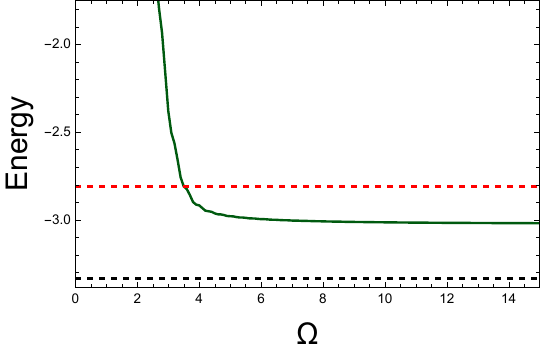}
    \caption{
The energy of the final-state obtained after implementing an adiabatic spiral as a function $\Omega$ for a comb of length 4 with $J=J_P=1$ and a switching time of $T=25$, starting from a N\'{e}el state. 
The spiral utilized the Hamiltonian given in Eq.~(\ref{eq:Hhatcomb}), with
$h_P(t)=0$ and $f(t)$ a linear function.
The black dashed line is the energy of the ground state and the red dashed line is the energy of the first excited state.}
    \label{fig:CombOmega}
\end{figure}

\noindent
To implement the adiabatic spiral in Eqs.~(\ref{eq:SpiralHam}) and (\ref{eq:spiraltilde}), 
specific choices have to be made for $h_P(t)$ and $f(t)$. 
As an example, we consider the Heisenberg comb that has recently been shown to 
reproduce the O(3) NL$\sigma$M in the continuum and infinite-volume limits.   
It has modest qubit requirements,  making it a good candidate for near term quantum simulations~\cite{PhysRevD.100.054505,PhysRevLett.126.172001}. 
The Hamiltonian is given by 
\begin{equation}
    \hat{H} = \sum_x  \left[ J \Vec{S}_{x,1} \cdot \Vec{S}_{x+1,1} + J_p \Vec{S}_{x,1} \cdot \Vec{S}_{x,2} \right]
    \ \ ,
    \label{eq:HeisenbergComb}
\end{equation}
where $\Vec{S}_{x,y}=\frac{1}{2} \Vec{\sigma}_{x,y}$ is the vector of Pauli matrices divided by $2$ at position $(x,y)$ on the lattice. 
An adiabatic spiral can be used to prepare the ground state of this model 
from a N\'{e}el state with the Hamiltonian
\begin{equation}
    \hat{H}(t) = \frac{1}{4} \sum_x \left[ J \hat{Z}_{x,1} \hat{Z}_{x+1,1} + J_p \hat{Z}_{x,1} \hat{Z}_{x,2} \right] + \sum_{x,y} \left[ \frac{\Omega}{2} \left(\frac{1}{\sqrt{3}}\hat{Z}_{x,y} + f(t) \hat{X}_{x,y} \right) + \frac{h_P(t)}{2} (-1)^{x+y} \hat{Z}_{x,y} \right] 
    \ \ \ .
    \label{eq:Hhatcomb}
\end{equation}
The simplest choices for $f(t)$ and $h_P(t)$ are linear functions of $t$. 
Once the functional forms of $f(t)$ and $h_P(t)$ have been chosen, 
implementing the adiabatic spiral further requires choices of $\Omega$, $T$, and $h_P(0)$. 
To ensure the eigenstates of the Ising model are as close to eigenstates of the Heisenberg model as possible, and the conditions of the adiabatic theorem are satisfied, 
$\Omega$ and $T$ should be taken to be as large as possible. 
As an example, Fig. \ref{fig:CombOmega} shows the energy of a state obtained using the adiabatic spiral as a function of $\Omega$ for a Heisenberg comb of length $4$ with $J=J_P=1$ and a switching time of $T=25$. 
In this calculation, $h_P(t)=0$ for all times 
and $f(t)$ was taken to be a linear function. Evolution under $\hat{H}(t)$ was evaluated numerically by computing $\prod_{n=1}^N e^{-i \frac{T}{N} \hat{H}(\frac{n}{N}T)}$ and increasing $N$ until convergence.
At small values of $\Omega$, the eigenstates of the Ising model and Heisenberg model are not close and the adiabatic spiral fails. 
At large $\Omega$, the adiabatic spiral is able to prepare a state with an energy below the energy of the first excited state. 
The reason the energy  of the state in Fig. \ref{fig:CombOmega} saturates above the ground-state energy is due to a combination of the amount of time used in the switching and the initial degeneracy in the ground state.

\begin{figure}[!ht]
    \centering
    \includegraphics[width=8.6cm]{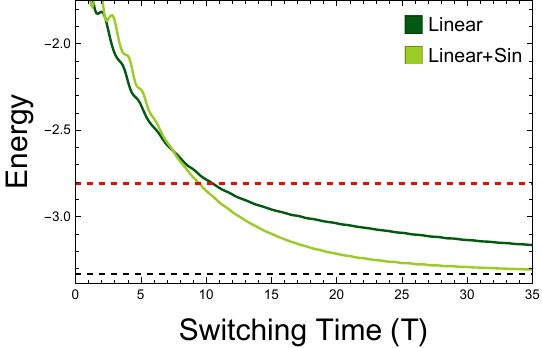}
    \caption{
    The dependence of the final state energy, after implementing an adiabatic spiral,
    on the total switching time used for a comb of length 4 with $J=J_P=1$. The dark green curve shows the energy obtained when $f(t)$ and $h_P(t)$ are taken to be linear functions. The light green curve shows the energy obtained when $f(t) = \sqrt{\frac{2}{3}} \left(\frac{t}{T} + \frac{1}{\pi} \sin(\pi \frac{t}{T})\right)$. The black dashed line is the energy of the ground state and the red dashed line is the energy of the first excited state.
    }
    \label{fig:CombPath}
\end{figure}

The energy of the state prepared by the adiabatic spiral can be lowered by taking a non-zero value of $h_P(0)$. 
Unlike with $\Omega$ and $T$, the energy of the state prepared by an adiabatic spiral is not monotonic in $h_P(0)$. 
If $h_P(0)$ is taken to be too large, 
the process of switching off the initial energy penalty in a finite amount of time can break adiabaticity, leading to a state with larger energy being prepared. 
An optimal value of $h_P(0)$ can be found variationally. 
Fig. \ref{fig:CombPath} shows the energy obtained from the adiabatic spiral 
with $\Omega = 8$ and $h_P(0) = 0.18$. 
This value for $h_P(0) = 0.18$ was selected by minimizing the energy obtained by performing the adiabatic spiral with $\Omega=8$ and $T=25$.

The adiabatic spiral's performance can be improved further by optimizing the path taken through parameter space. 
If $h_p(t)$ is taken to be a linear function of time, and $f(t)$ taken to be
\begin{equation}
    f(t) = \sqrt{\frac{2}{3}} \left(\frac{t}{T} + \sum_{n<N} \beta_n \sin\left(n \pi \frac{t}{T}\right) \right) 
    \  \ \ ,
    \label{eq:SwitchParameters}
\end{equation}
an optimal path through parameter space can be found by 
minimizing the energy obtained as a function of the $\beta_n$s~\cite{https://doi.org/10.48550/arxiv.2203.01948}. 
Note that if the maximum driving field strength is limited on the analog simulator, this will constrain the values that the $\beta_n$s can take. For long switching times, assuming that the maximum value of the driving field is obtained at the end of the spiral  and truncating at $\beta_1$, it was found that the energy obtained by the adiabatic spiral was monotonic as a function of $\beta_1$. 
This means that in practice, the largest value of $\beta_1$ allowed by the constraint should be used so that the maximum value of the driving field is obtained at the 
end of the spiral, which is $\beta_1 = \frac{1}{\pi}$. 
The energy of the state obtained by performing the adiabatic spiral with this path is also 
shown in Fig.~\ref{fig:CombPath}. 
At long switching times, 
this path can offer an improved performance over linear switching. 
It was also found that including more terms in the expansion in 
Eq.~(\ref{eq:SwitchParameters}) 
offered minimal improvements in the performance of the adiabatic spiral.

%%%%%%%%%%%%%%%%%%%%%%%%%%%%%%%%%%%%%%%%%%%%%%%%%%%%%%%
\section{Potential Hardware Implementations}
\subsection{Rydberg Atoms}
\begin{figure}[!ht]
    \centering
    \includegraphics[width=8.6cm]{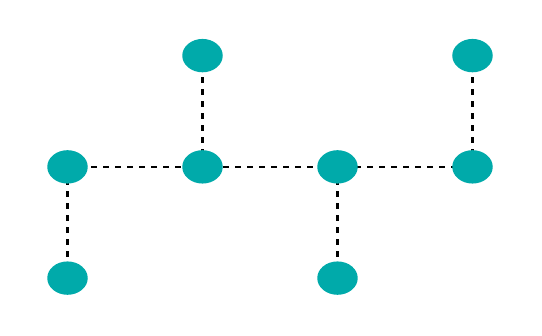}
    \caption{An arrangement of Rydberg atoms that can be used to perform an analog quantum simulation of the Heisenberg comb.}
    \label{fig:CombArray}
\end{figure}
\noindent
Arrays of Rydberg atoms are an experimental platform that could be used to implement the adiabatic spiral. If the atoms in the Rydberg state interact through the Van der Waals interaction, the Hamiltonian describing their dynamics is
\begin{equation}
    \hat{H}^{\text{Ryd.}}(t) = \frac{\Omega(t)}{2} \sum_{i} \hat{X}_i - \Delta(t) \sum_{i} \hat{n}_i + \sum_{i<j}\frac{V_0}{\abs{\Vec{x}_i - \Vec{x}_j}^6} \hat{n}_i \hat{n}_j
    \ \ ,
\end{equation}
where $\hat{n}_i$ is the Rydberg state occupation at site $i$, $\Vec{x}_i$ is the position of site $i$ and $\hat{X}_i$ couples the ground state of the atom at site $i$ to its Rydberg state. This native implementation of Ising interactions has enabled the use of Rydberg atoms to perform analog simulations of the Ising model in 1D and 2D~\cite{saffman2010quantum,bernien2017probing,PhysRevX.8.021070,PhysRevX.8.021069,kim2018detailed,de2019observation,keesling2019quantum,morgado2021quantum,semeghini2021probing,ebadi2021quantum}. 
Their native Ising interactions allows for the implementation of adiabatic spirals.

This is not the only technique that can be used to simulate the Heisenberg interaction on Rydberg atoms. 
In recent work, a method of simulating Heisenberg interactions on Rydberg atoms making use of  dipole-dipole interactions and an external microwave field was proposed~\cite{PRXQuantum.3.020303}. 
This approach uses global microwave pulses to rotate the dipole-dipole interaction to generate time evolution approximating an XXZ Heisenberg Hamiltonian, analogous to how one could Trotterize the Heisenberg interaction on a digital quantum computer. While this proposal makes use of dipole-dipole interactions, the same approach could be used for Rydberg atom simulations that use the Van der Waals interaction to couple atoms. In that experimental setup, an effective $\hat{Z}_i \hat{Z}_j$ is available. In principle, the ground state of a Heisenberg model could be prepared on this hardware using the adiabatic spiral or by using global pulses to perform a Trotterized adiabatic switching. The choice of which method to use will depend on the specific analog simulator being used. As a case study, we can consider hardware parameters used in a recent experiment to study phases of the 2D Ising model~\cite{ebadi2021quantum}. This experimental setup used a driving field with a max amplitude of $\Omega_{max}=2\pi \times 4.3 \text{MHz}$ and had a max coherence time of $3 \mu\text{s}$. With these hardware parameters, the preparation of the ground state of a Heisenberg comb with the adiabatic spiral parameters from the previous section can be performed by placing atoms in an array as shown in Fig. \ref{fig:CombArray} with a lattice spacing of $a\approx10.5\mu\text{m}$. Note that the Hamiltonian simulated will not quite be the same as the Hamiltonian from the previous section due to the $\frac{1}{r^6}$ coupling between the atoms, but the alternating vertical positions of atoms will ensure the next-to-nearest neighbor interactions will be suppressed. This same arrangement of atoms could be used to implemented Trotterized adiabatics. In this approach, $\frac{\pi}{2}$ pulses are alternated with periods of the external field turned off to construct a first order Trotterized approximation of time evolution according to
\begin{equation} \label{eq:LinearAdiabatic}
    \hat{H}(t) = \sum_{i,j} J_{i,j} \left(\hat{Z}_i \hat{Z}_j + \frac{t}{T} \left(\hat{X}_i \hat{X}_j + \hat{Y}_i \hat{Y}_j \right) \right) + h_P \left(1 - \frac{t}{T}\right)\sum_i (-1)^i \hat{Z}_i \ \ \ ,
\end{equation}
where $(-1)^i$ represents an alternating pattern of signs that breaks the initial degeneracy of the Hamiltonian. The errors in this state preparation method come from a finite switching time, standard Trotterization errors and the amount of time required to implement the $\frac{\pi}{2}$ pulses. This is similar in spirit to previous proposals to simulate time evolution using global controls~\cite{masanes2002time}. The explicit pulse sequences for first order Trotterization can be found in Appendix~\ref{app:RydbergTrotter}.

Figure~\ref{fig:TrotterVSpiral} shows the energy that can be obtained using the path optimized adiabatic spiral and Trotterized adiabatics with different numbers of Trotter steps as a function of total runtime on the analog simulator. At short times, a small number of Trotter steps are able to outperform the spiral. However, if the number of Trotter steps is fixed and the switching time is increased, the Trotterization error will grow and the energy of the state obtained eventually will increase. This can be mitigated by performing more Trotter steps, but as the number of Trotter steps increases the contribution to the error from the finite pulse time will accumulate. As shown in Fig. \ref{fig:TrotterVSpiral}, this effectively puts a limit on the total number of Trotter steps that can be performed before adiabatics is no longer being effectively simulated. As shown in Appendix~\ref{app:ImprovedTrotter}, some of these errors can be cancelled at leading order with a modified pulse sequence. The dashed curves in Fig.~\ref{fig:TrotterVSpiral} show the energy of the state that can be obtained by mitigating these errors at leading order in Trotterized adiabatics. The adiabatic spiral does not suffer from either of these issues. Increasing the amount of time used in the adiabatic spiral only increases the overlap with the ground state. It should also be noted that in the Trotterization simulations it was assumed that the drive field can be instantly ramped to its maximum value. In practice, this is not possible and the shape of the pulse while ramping on the drive field will contribute to errors in the time evolution. Given these limitations, it appears that the adiabatic spiral is a better choice for preparing the ground state of a Heisenberg model than Trotterized adiabatics on a Rydberg system. It should also be noted that the pulse sequences required to implement the adiabatic spiral are quite similar to other pulse sequences that have been implemented on Rydberg atoms~\cite{bernien2017probing,keesling2019quantum,semeghini2021probing,ebadi2021quantum}. As such, it is expected that the adiabatic spiral will be robust against experimental errors when implemented on Rydberg atoms. 
\begin{figure}
    \centering
    \includegraphics[width=8.6cm]{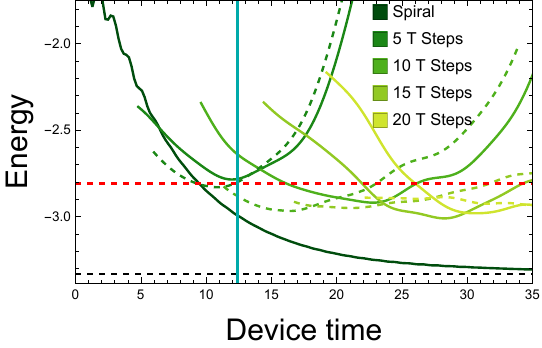}
    \caption{
The dependence of the final-state energy from an adiabatic spiral and Trotterized adiabatics on the total switching time used for a comb of length 4 with $J=J_P=1$. The solid lines used the Trotter sequence from Appendix~\ref{app:RydbergTrotter} and the dashed curves used the improved Trotter sequence from Appendix~\ref{app:ImprovedTrotter}. The horizontal black dashed line is the energy of the ground state and the horizontal red dashed line is the energy of the first excited state. The vertical blue line indicates the coherence time of $3 \mu \text{s}$ in units of $J^{-1}$.}
    \label{fig:TrotterVSpiral}
\end{figure}

%%%%%%%%%%%%%%%%%%%%%%%%%%%%%%%%%%%%
\subsection{D-Wave}
\noindent
Another experimental platform that could potentially implement the adiabatic spiral is D-Wave's quantum annealer. 
The D-Wave quantum annealer implements time evolution according to the Hamiltonian,
\begin{equation}
    \hat{H}^{\text{DW}}(t) = -\frac{\text{A}(s(t))}{2} \sum_i \hat{X}_i + \frac{\text{B}(s(t))}{2} \left(\sum_i h_i \hat{Z}_i + \sum_{i<j} J_{ij} \hat{Z}_i \hat{Z}_j \right)
    \ \ ,
\end{equation}
where $\text{A}(s)$ and $\text{B}(s)$ are fixed functions, but $s(t)$, $h_i$ and $J_{ij}$ can be programmed by the user, up to some constraints. 
Although it is designed to solve optimization problems, 
D-Wave's hardware has been used to perform analog simulations of the Ising model~\cite{gardas2018defects,doi:10.1126/science.aat2025,King_2018,PhysRevLett.124.090502,King_2021,PhysRevResearch.2.033369,PRXQuantum.1.020320,https://doi.org/10.48550/arxiv.2003.14244,PhysRevA.102.042403,doi:10.1126/science.abe2824,https://doi.org/10.48550/arxiv.2202.05847}. 
In previous work, D-Wave's quantum annealer has also been used to simulate time evolution and perform state preparation for lattice gauge theories, however these mappings to quantum simulation have more in common with digital quantum simulation than the analog protocol we are proposing~\cite{ARahman:2021ktn,Illa:2022jqb,Farrell:2022wyt}. 
To implement the adiabatic spiral on the D-Wave annealer, 
a set of qubits with a comb pattern coupling must be selected. 
Given the connectivity of the hardware, this is straightforward to do. 
$J_{ij}$ and $h_{i}$ should be chosen so that the implemented Hamiltonian is
\begin{equation}
    \hat{H}^{\text{DW}}(t) = -\frac{\text{A}(s(t))}{2} \sum_{x,y} \hat{X}_{x,y} + \frac{\text{B}(s(t))}{2} \left(\sum_{x,y} h \hat{Z}_{x,y} + \sum_{x} J  \hat{Z}_{x,1} \hat{Z}_{x+1,1} + J_p  \hat{Z}_{x,1} \hat{Z}_{x,2} \right) \ \ \ .
\end{equation}
\begin{figure}[!ht]
\centering
    \includegraphics[width=8.6cm]{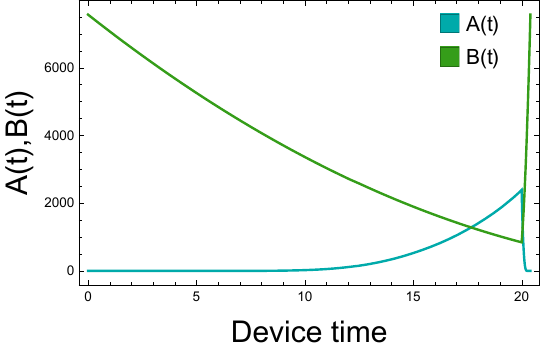}
    \caption{Annealing schedule on the D-Wave's {\tt Advantage} 6.1 to implement the adiabatic spiral with $h=2$.  
    $\text{A}(t)$ and $\text{B}(t)$ are given in units of MHz and time is in units of $\mu$s.
    }
    \label{fig:DWaveSchedule}
\end{figure}
\begin{figure}
\centering
    \begin{tikzpicture}
    \node(a){\includegraphics[width=8.6cm]{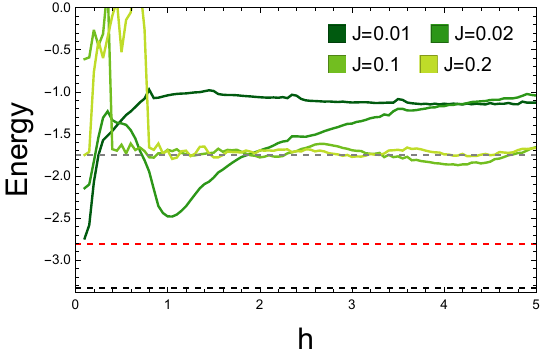}};
    \node at (a.north west) [anchor=north,xshift=2mm,yshift=0mm]
    {\includegraphics[width=0.04\textwidth]{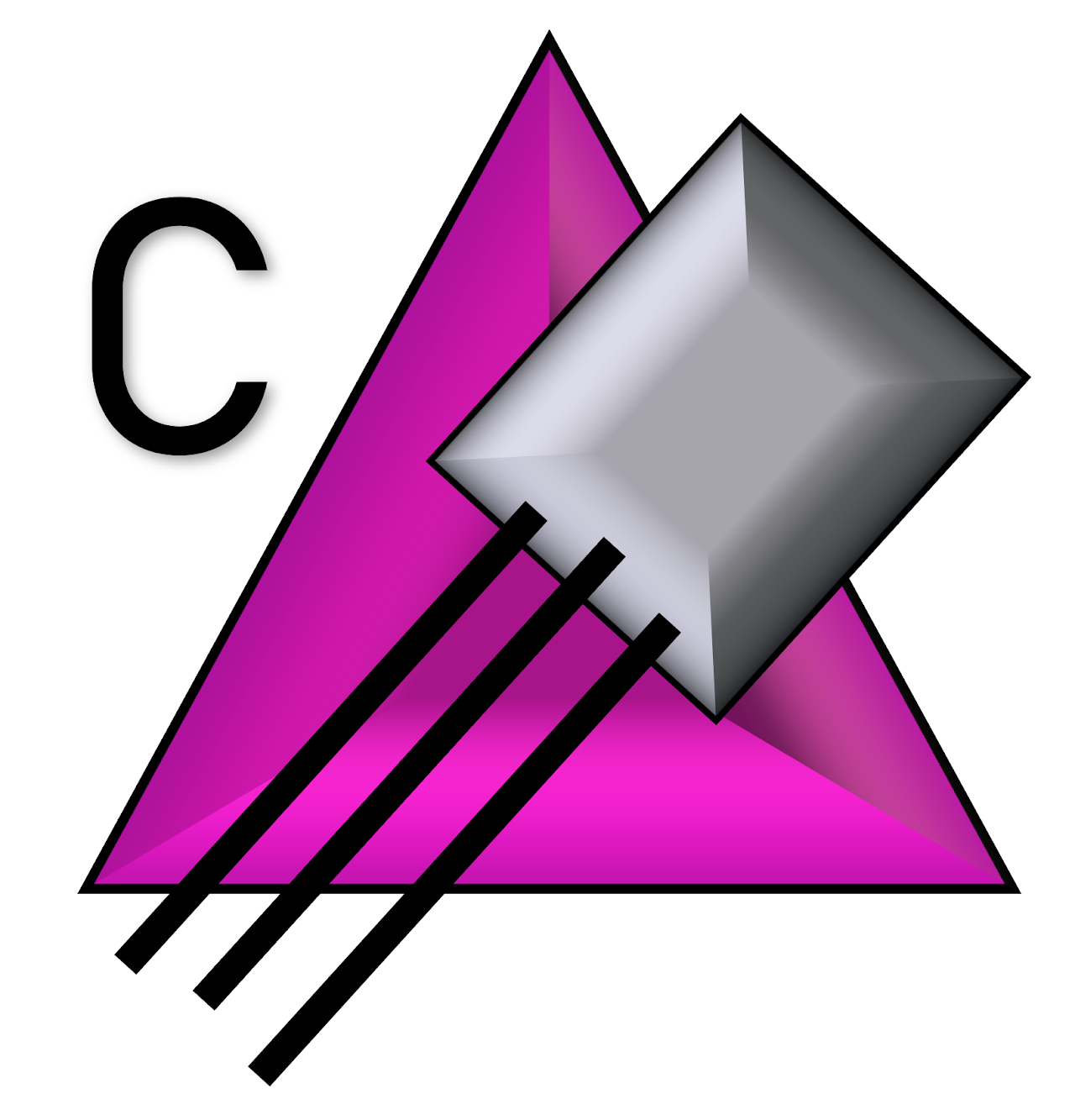}};
    \end{tikzpicture}
    \caption{
    Energy of the state produced on a simulation of D-Wave's {\tt Advantage} 6.1 hardware 
     for different values of $J$ and $h$. The gray line is the energy of the initial state. The black line is the vacuum energy, and the red line is the energy of the first excited state. Note that the energy shown here is measured in units of $J$.}
    \label{fig:DWavePaths}
\end{figure}
\begin{figure}
\centering
    \begin{tikzpicture}
    \node(a){\includegraphics[width=8.6cm]{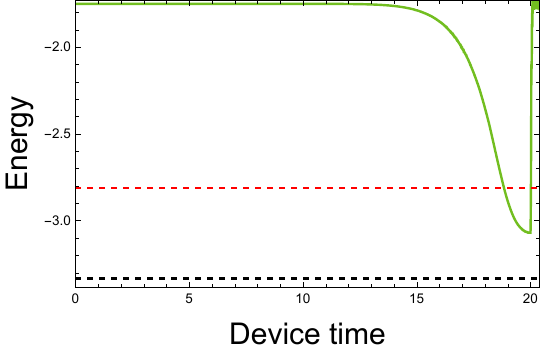}};
    \node at (a.north west) [anchor=north,xshift=2mm,yshift=0mm]
    {\includegraphics[width=0.04\textwidth]{fig/ideal-quantum-simulation-classical-hardware.png}};
    \end{tikzpicture}
    \caption{
    Energy of the state produced on a simulation of D-Wave's {\tt Advantage} 6.1
    as a function of time for a comb of length 4 with $J=J_P=0.1$ and $h=2$. 
    The black dashed line is the energy of the ground state 
    and the red dashed line is the energy of the first excited state. 
    Note that the energy is given in units of $J$.}
    \label{fig:DWaveEnergy}
\end{figure}
The spiral can be implemented by using D-Wave's reverse annealing function to begin in a N\'{e}el state with 
$\text{A}(s(0)) = 0$, 
and adiabatically evolve to $s^*$ such that $\text{A}(s^*)=\sqrt{2}\text{B}(s^*) h$. 
As long as $J,J_p \ll h$, 
the adiabatic spiral will be able to prepare an equal superposition of the ground state and the first excited state of the Heisenberg comb. 
The pure ground state will not be prepared because an energy penalty cannot be implemented to break the degeneracy within the initial state.
Note that to perform a measurement on the D-Wave QPU, the pulse schedule must end with $\text{A}(s=1)=0$. 
The process of turning off $\text{A}(s)$ will create errors in the state preparation whose size are controlled by the slew rate of $\text{A}(s)$ and the choice of $s^*$. 
One choice of time dependence of $\text{A}(t)$ and $\text{B}(t)$ to implement the adiabatic spiral 
is shown in Fig.~\ref{fig:DWaveSchedule}. 
As an explicit example, we once again consider the Heisenberg comb of length $4$ with $J=J_p$. The energy of states prepared with classical simulations 
({\tt Mathematica} and {\tt python})
of the D-Wave {\tt Advantage} 6.1 hardware for different choices of $J$ and $h$ 
are shown in Fig.~\ref{fig:DWavePaths}. 
For small values of $J$ and $h$, 
simulations of the D-Wave hardware are able to produce states 
with a lower energy than the initial state. 
However, the quantum device has calibration errors on the order of 
$\delta J \approx 4*10^{-3}$,  making it unlikely that these parameters 
can be used in practice. 
For larger values of $J$, the simulations  cannot reach states with 
energy lower than the initial state. 
As an example, the energy of states as a function of time for $J=0.1$ and $h=2$ 
found in simulations of {\tt Advantage} are shown in Fig. \ref{fig:DWaveEnergy}. 
As seen, {\tt Advantage} is expected to be able to prepare a low energy state 
with respect to the Heisenberg Hamiltonian part-way through a simulation. 
However, in the process of turning off $\text{A}(s)$ as required to perform a measurement, 
the time evolution is still approximately adiabatic, 
and the system ends up in a final state with large overlap with the initial state. 
If the slew rate of {\tt Advantage} could be increased, or if measurements could be performed without fully turning off $\text{A}(t)$, 
it would be possible to perform measurements of the 
low-energy state prepared by the adiabatic spiral 
and potentially use {\tt Advantage} in a scalable manner 
for quantum simulation of the Heisenberg model. Implementing the adiabatic spiral on {\tt Advantage} also suffers from the complication that the single qubit terms must be tuned to a specific value to successfully implement the spiral. While a pulse schedule is provided by D-wave for this hardware, the pulses actually implemented can vary from this schedule by up to 30\%. This is not an issue for the classical optimization problems the {\tt Advantage} quantum annealer is designed to solve, but represents a significant issue for reliable implementation of the adiabatic spiral on this hardware.

%%%%%%%%%%%%%%%%%
\section{Discussion}
\noindent
In this work, a method of simulating the Heisenberg model on quantum hardware with native Ising interactions
was extended to perform adiabatic state preparation of the Heisenberg ground state. 
Due to the number of quantum simulation platforms able to natively implement the Ising model, this technique is likely to have experimental applicability. 
It was shown that it should be feasible to implement this technique on Rydberg atoms, and that it can out-perform a Trotterized approach to  adiabatic state preparation. The feasibility of implementing the adiabatic spiral on D-Wave quantum annealers was also investigated, 
and it was found that the requirement of turning off the $\hat{X}$ driving field at the end of the calculation limits its efficacy. 
The ability of the adiabatic spiral to prepare low-energy states of the Heisenberg model on these hardware platforms suggests that near-term quantum simulators 
can be used to perform analog quantum simulations of the $O(3)$ NL$\sigma$M, and other such theories of importance to nuclear and particle physics. While the case of the fully symmetric XXX Heisenberg model was focused on in this work, the adiabatic spiral can also be applied to the XXZ Heisenberg model as well.
As the approximation of the dynamics of the Heisenberg model by an Ising model with strong drive fields is a special case of the more generic phenomena of prethermalization~\cite{abanin2017rigorous,PhysRevX.7.011026,PhysRevB.100.020301,https://doi.org/10.48550/arxiv.2103.07485},
similar ideas may find applicability for other Hamiltonians on analog quantum simulators 
with different native couplings.

%%%%%%%%%%%%%%%%%%%%%%%%%
\begin{acknowledgments}
The authors would like to acknowledge Pavel Lougovski, Peter Komar and Cedric Lin from the Amazon Braket team for contributions to developing this technique. The views expressed are those of the authors and do not reflect the official policy or position of AWS. The authors would also like to acknowledge many useful conversations with Hersh Singh. The material presented here was funded 
in part by the DOE QuantISED program through the theory  consortium ``Intersections of QIS and Theoretical Particle Physics'' at Fermilab with Fermilab Subcontract No. 666484,
and in part by U.S. Department of Energy, Office of Science, Office of Nuclear Physics, Inqubator for Quantum Simulation (IQuS) under Award Number DOE (NP) Award DE-SC0020970 and the Quantum Science Center (QSC) a National Quantum Information
Science Research Center of the U.S. Department of Energy.
\end{acknowledgments}

%%%%%%%%%%%%%%%%%%%%%%%%
\clearpage
\appendix

%%%%%%%%%%%%%%
\section{First Order Trotterization with Rydberg Atoms}
\label{app:RydbergTrotter}
\noindent
In this section, it will be shown how to engineer a Trotterized approximation to an adiabatic turning on of the Heisenberg interaction on an array of Rydberg atoms. Assuming access to a global and staggered driving field, the Hamiltonian describing the time evolution of an array of Rydberg atoms is given by 
\begin{equation} \label{eq:IsingTrotterResource}
\hat{H}^{\text{Ryd.}}(t) = \sum_{i,j} J_{i,j} \hat{Z}_i \hat{Z}_{j} + \sum_i \vec{\omega}(t) \cdot \vec{S}_i +\sum_i \frac{h(t)}{2} (-1)^i \hat{Z}_{i} 
\ \ \ .
\end{equation}
It will also be assumed that the strength of the global drive field is limited, $|\vec{\omega}(t)|\leq \Omega$, as is the case in actual Rydberg atom experiments. The approximation to Heisenberg time evolution will be assembled from the following set of global analog gates,
\begin{align} \label{eq:AnalogGateSet}
    R^\pm_X(\epsilon) & = \exp{-i\epsilon \sum_{i,j} J_{i,j} \hat{Z}_i \hat{Z}_j \mp \frac{i \pi}{4} \sum_j \hat{X}_j} \ \ \ , \nonumber \\
    R^\pm_Y(\epsilon) & = \exp{-i \epsilon \sum_{i,j} J_{i,j} \hat{Z}_i \hat{Z}_j \mp \frac{i \pi}{4} \sum_j \hat{Y}_j} \ \ \ ,  \nonumber \\
    R_Z(t,\kappa) & = \exp{-i t \sum_{i,j} J_{i,j} \hat{Z}_i \hat{Z}_j  - i \frac{\kappa}{2} \sum_j (-1)^j \hat{Z}_j} \ \ \ .
\end{align}
$R^\pm_X(\epsilon)$ and $R^\pm_Y(\epsilon)$ correspond to global $\frac{\pi}{2}$-pulses about the x and y axes respectively and can be generated using the global drive field. $R_Z(t,\kappa)$ can be generated by only turning on the staggered driving field. Due to the maximum driving field strength, $\Omega$, the $\frac{\pi}{2}$-pulses, $R^\pm_{X,Y}$, need a minimum device time $\epsilon = \frac{\pi}{2\Omega}$ 
during which the $\hat{Z}_i\hat{Z}_j$ interaction cannot be ``turned off.'' This leads to ${\cal O}(\epsilon)$ cross-talk errors that will be studied in further detail in Appendix~\ref{app:ImprovedTrotter}. With this analog gate set, a first order Trotter approximation to a generic XXZ Heisenberg evolution with a staggered single qubit term is given by
\begin{align}
    U_{XXZ}(x,z, h) & = \exp{-i \sum_{i,j} J_{i,j} \left(x\hat{X}_i \hat{X}_j + x\hat{Y}_i \hat{Y}_j + z \hat{Z}_i \hat{Z}_j \right) - i h \sum_i (-1)^i \hat{Z}_{i} } \nonumber \\
    & \approx \left(\prod_j \hat{Z}_j\right) R_Z(z,h) R^+_X(\epsilon) R_Z(x,0) R^+_X(\epsilon) R^+_Y(\epsilon) R_Z(x,0) R^+_Y(\epsilon) \ \ \ .
    \label{eq;NaiveTrotter1}
\end{align}
Note that since $\sum \hat{Z}_j$ commutes with the XXZ Heisenberg Hamiltonian, the product over $\hat{Z}_j$'s can be neglected when performing adiabatic state preparation as its only contribution is to change the overall phase of the prepared eigenstate. With this pulse sequence, the time evolution generated by the Hamiltonian in Eq.~\ref{eq:LinearAdiabatic} can be approximated by
\begin{equation}
    \mathcal{T}e^{-i\int^T_0 dt \ \hat{H}(t)} \approx \prod_{n=1}^N R_Z\left(\frac{T}{N},\frac{T}{N} h_P \left(1 - \frac{n}{N}\right)\right) R^+_X(\epsilon) R_Z\left(\frac{Tn}{N^2},0\right) R^+_X(\epsilon) R^+_Y(\epsilon) R_Z\left(\frac{Tn}{N^2},0\right) R^+_Y(\epsilon) \ \ \ .
\end{equation}

%%%%%%%%%%%%%%
\section{Cross-talk Mitigated Trotter Sequence}
\label{app:ImprovedTrotter}
\noindent
The accuracy of the Trotterized approximation to the time-evolution operator can be improved by using shorter Trotter steps or higher order formulas. However, for Trotter step sizes comparable to the $\frac{\pi}{2}$ pulse length, it becomes important to compensate for the ${\cal O}(\epsilon)$ contributions to the error. In this section, it will be shown how a modification of the pulse sequence in Appendix~\ref{app:RydbergTrotter} can be performed to cancel this error at leading order. To leading order in $\epsilon$, the $\frac{\pi}{2}$ rotations are given by
\begin{align}
   R^\pm_X(\epsilon) & = \exp{ \mp \frac{i \pi}{4} \sum_j \hat{X}_j}\exp{-\frac{i}{2 \Omega} \sum_{i,j} J_{i,j} \left( \frac{\pi}{2}\left(\hat{Z}_i \hat{Z}_j + \hat{Y}_i \hat{Y}_j \right) \pm \left( \hat{Z}_i \hat{Y}_j + \hat{Y}_i \hat{Z}_j\right) \right) }  + {\cal O}(\epsilon^2) \ \ \ , \nonumber \\
   R^\pm_Y(\epsilon) & = \exp{ \mp \frac{i \pi}{4} \sum_j \hat{Y}_j}\exp{-\frac{i}{2 \Omega} \sum_{i,j} J_{i,j} \left( \frac{\pi}{2}\left(\hat{Z}_i \hat{Z}_j + \hat{X}_i \hat{X}_j \right) \mp \left( \hat{Z}_i \hat{X}_j + \hat{X}_i \hat{Z}_j\right) \right) }+ {\cal O}(\epsilon^2) \ \ \ .
\end{align}
Using these expressions, it can be shown that at leading order in $\epsilon$ the pulse sequence from Appendix~\ref{app:RydbergTrotter} is
\begin{align}
     & R_Z(z,h) R^+_X(\epsilon) R_Z(x,0) R^+_X(\epsilon) R^+_Y(\epsilon) R_Z(x,0) R^+_Y(\epsilon) \approx \left(\prod_j \hat{Z}_j\right) \nonumber \\ 
     & \exp{-i \sum_{i,j} J_{i,j} \left( (x+\epsilon)\hat{X}_i \hat{X}_j + (x+\epsilon)\hat{Y}_i \hat{Y}_j + (z+2\epsilon) \hat{Z}_i \hat{Z}_j \right) - i h \sum_i (-1)^i \hat{Z}_{i} }  \ \ \ .
\end{align}
By shifting the length of pulses in the sequence from Appendix~\ref{app:RydbergTrotter}, the $\mathcal{O}(\epsilon)$ terms can be absorbed into the time evolution. This motivates the improved adiabatic Trotter sequence given by 
\begin{align}
        \mathcal{T}e^{-i\int^T_0 dt \ \hat{H}(t)} \approx \prod_{n=1}^N & R_Z\left(\frac{T}{N}-2\epsilon,\frac{T}{N} h_P \left(1 - \frac{n}{N}\right)\right) \nonumber\\ 
        & R^+_X(\epsilon) R_Z\left( \left(\frac{Tn}{N^2}-\epsilon\right) \theta\left(\frac{Tn}{N^2}-\epsilon\right),0\right) R^+_X(\epsilon) \nonumber \\ 
        & R^+_Y(\epsilon) R_Z\left(\left(\frac{Tn}{N^2}-\epsilon\right) \theta\left(\frac{Tn}{N^2}-\epsilon\right),0\right) R^+_Y(\epsilon) \ \ \ ,
\end{align}
where $\theta(x)$ is the Heaviside theta function. When $\frac{T}{N^2}$ is larger than $\epsilon$, this pulse sequence cancels the cross-talk errors at $\mathcal{O}(\epsilon)$. When the Trotter step size is taken to be smaller, this pulse sequence will only cancel the cross-talk errors for the later steps in the sequence.

\bibliographystyle{quantum}
\bibliography{ref,bibi,biblioKRS}

\providecommand{\noopsort}[1]{}\providecommand{\singleletter}[1]{#1}%
\begin{thebibliography}{100}

\bibitem{Zohar:2011cw}
Erez Zohar and Benni Reznik.
\newblock ``{Confinement and lattice QED electric flux-tubes simulated with
  ultracold atoms}''.
\newblock \href{https://dx.doi.org/10.1103/PhysRevLett.107.275301}{Phys. Rev.
  Lett. {\bf 107}, 275301}~(2011).
\newblock  \href{http://arxiv.org/abs/1108.1562}{arXiv:1108.1562}.

\bibitem{Zohar:2012ay}
Erez Zohar, J.~Ignacio Cirac, and Benni Reznik.
\newblock ``{Simulating Compact Quantum Electrodynamics with ultracold atoms:
  Probing confinement and nonperturbative effects}''.
\newblock \href{https://dx.doi.org/10.1103/PhysRevLett.109.125302}{Phys. Rev.
  Lett. {\bf 109}, 125302}~(2012).
\newblock  \href{http://arxiv.org/abs/1204.6574}{arXiv:1204.6574}.

\bibitem{Tagliacozzo:2012vg}
L.~Tagliacozzo, A.~Celi, A.~Zamora, and M.~Lewenstein.
\newblock ``{Optical Abelian Lattice Gauge Theories}''.
\newblock \href{https://dx.doi.org/10.1016/j.aop.2012.11.009}{Annals Phys. {\bf
  330}, 160--191}~(2013).
\newblock  \href{http://arxiv.org/abs/1205.0496}{arXiv:1205.0496}.

\bibitem{Zohar:2012ts}
Erez Zohar, J.~Ignacio Cirac, and Benni Reznik.
\newblock ``{Simulating (2+1)-Dimensional Lattice QED with Dynamical Matter
  Using Ultracold Atoms}''.
\newblock \href{https://dx.doi.org/10.1103/PhysRevLett.110.055302}{Phys. Rev.
  Lett. {\bf 110}, 055302}~(2013).
\newblock  \href{http://arxiv.org/abs/1208.4299}{arXiv:1208.4299}.

\bibitem{PhysRevLett.110.125303}
D.~Banerjee, M.~B\"ogli, M.~Dalmonte, E.~Rico, P.~Stebler, U.-J. Wiese, and
  P.~Zoller.
\newblock ``{Atomic Quantum Simulation of $\mathbf{U}(N)$ and $\mathrm{SU}(N)$
  Non-Abelian Lattice Gauge Theories}''.
\newblock \href{https://dx.doi.org/10.1103/PhysRevLett.110.125303}{Phys. Rev.
  Lett. {\bf 110}, 125303}~(2013).
\newblock  \href{http://arxiv.org/abs/1211.2242}{arXiv:1211.2242}.

\bibitem{Hauke:2013jga}
Philipp Hauke, David Marcos, Marcello Dalmonte, and Peter Zoller.
\newblock ``{Quantum simulation of a lattice Schwinger model in a chain of
  trapped ions}''.
\newblock \href{https://dx.doi.org/10.1103/PhysRevX.3.041018}{Phys. Rev. X {\bf
  3}, 041018}~(2013).
\newblock  \href{http://arxiv.org/abs/1306.2162}{arXiv:1306.2162}.

\bibitem{Marcos:2014lda}
D.~Marcos, P.~Widmer, E.~Rico, M.~Hafezi, P.~Rabl, U.~J. Wiese, and P.~Zoller.
\newblock ``{Two-dimensional Lattice Gauge Theories with Superconducting
  Quantum Circuits}''.
\newblock \href{https://dx.doi.org/10.1016/j.aop.2014.09.011}{Annals Phys. {\bf
  351}, 634--654}~(2014).
\newblock  \href{http://arxiv.org/abs/1407.6066}{arXiv:1407.6066}.

\bibitem{Kuno:2014npa}
Yoshihito Kuno, Kenichi Kasamatsu, Yoshiro Takahashi, Ikuo Ichinose, and Tetsuo
  Matsui.
\newblock ``{Real-time dynamics and proposal for feasible experiments of
  lattice gauge\textendash{}Higgs model simulated by cold atoms}''.
\newblock \href{https://dx.doi.org/10.1088/1367-2630/17/6/063005}{New J. Phys.
  {\bf 17}, 063005}~(2015).
\newblock  \href{http://arxiv.org/abs/1412.7605}{arXiv:1412.7605}.

\bibitem{Bazavov:2015kka}
Alexei Bazavov, Yannick Meurice, Shan-Wen Tsai, Judah Unmuth-Yockey, and Jin
  Zhang.
\newblock ``{Gauge-invariant implementation of the Abelian Higgs model on
  optical lattices}''.
\newblock \href{https://dx.doi.org/10.1103/PhysRevD.92.076003}{Phys. Rev. D
  {\bf 92}, 076003}~(2015).
\newblock  \href{http://arxiv.org/abs/1503.08354}{arXiv:1503.08354}.

\bibitem{Kasper:2015cca}
V.~Kasper, F.~Hebenstreit, M.~Oberthaler, and J.~Berges.
\newblock ``{Schwinger pair production with ultracold atoms}''.
\newblock \href{https://dx.doi.org/10.1016/j.physletb.2016.07.036}{Phys. Lett.
  B {\bf 760}, 742--746}~(2016).
\newblock  \href{http://arxiv.org/abs/1506.01238}{arXiv:1506.01238}.

\bibitem{Brennen:2015pgn}
G.~K. Brennen, G.~Pupillo, E.~Rico, T.~M. Stace, and D.~Vodola.
\newblock ``{Loops and Strings in a Superconducting Lattice Gauge Simulator}''.
\newblock \href{https://dx.doi.org/10.1103/PhysRevLett.117.240504}{Phys. Rev.
  Lett. {\bf 117}, 240504}~(2016).
\newblock  \href{http://arxiv.org/abs/1512.06565}{arXiv:1512.06565}.

\bibitem{Kuno:2016xbf}
Yoshihito Kuno, Shinya Sakane, Kenichi Kasamatsu, Ikuo Ichinose, and Tetsuo
  Matsui.
\newblock ``{Atomic quantum simulation of a three-dimensional U(1) gauge-Higgs
  model}''.
\newblock \href{https://dx.doi.org/10.1103/PhysRevA.94.063641}{Phys. Rev. A
  {\bf 94}, 063641}~(2016).
\newblock  \href{http://arxiv.org/abs/1605.02502}{arXiv:1605.02502}.

\bibitem{Martinez:2016yna}
Esteban~A. Martinez, Christine~A. Muschik, Philipp Schindler, Daniel Nigg,
  Alexander Erhard, Markus Heyl, Philipp Hauke, Marcello Dalmonte, Thomas Monz,
  Peter Zoller, and Rainer Blatt.
\newblock ``{Real-time dynamics of lattice gauge theories with a few-qubit
  quantum computer}''.
\newblock \href{https://dx.doi.org/10.1038/nature18318}{Nature {\bf 534},
  516--519}~(2016).
\newblock  \href{http://arxiv.org/abs/1605.04570}{arXiv:1605.04570}.

\bibitem{Kasper:2016mzj}
V.~Kasper, F.~Hebenstreit, F.~Jendrzejewski, M.~K. Oberthaler, and J.~Berges.
\newblock ``{Implementing quantum electrodynamics with ultracold atomic
  systems}''.
\newblock \href{https://dx.doi.org/10.1088/1367-2630/aa54e0}{New J. Phys. {\bf
  19}, 023030}~(2017).
\newblock  \href{http://arxiv.org/abs/1608.03480}{arXiv:1608.03480}.

\bibitem{Muschik:2016tws}
Christine Muschik, Markus Heyl, Esteban Martinez, Thomas Monz, Philipp
  Schindler, Berit Vogell, Marcello Dalmonte, Philipp Hauke, Rainer Blatt, and
  Peter Zoller.
\newblock ``{U(1) Wilson lattice gauge theories in digital quantum
  simulators}''.
\newblock \href{https://dx.doi.org/10.1088/1367-2630/aa89ab}{New J. Phys. {\bf
  19}, 103020}~(2017).
\newblock  \href{http://arxiv.org/abs/1612.08653}{arXiv:1612.08653}.

\bibitem{Gonzalez-Cuadra:2017lvz}
Daniel Gonz\'alez-Cuadra, Erez Zohar, and J.~Ignacio Cirac.
\newblock ``{Quantum Simulation of the Abelian-Higgs Lattice Gauge Theory with
  Ultracold Atoms}''.
\newblock \href{https://dx.doi.org/10.1088/1367-2630/aa6f37}{New J. Phys. {\bf
  19}, 063038}~(2017).
\newblock  \href{http://arxiv.org/abs/1702.05492}{arXiv:1702.05492}.

\bibitem{Klco:2018kyo}
N.~Klco, E.~F. Dumitrescu, A.~J. McCaskey, T.~D. Morris, R.~C. Pooser, M.~Sanz,
  E.~Solano, P.~Lougovski, and M.~J. Savage.
\newblock ``{Quantum-classical computation of Schwinger model dynamics using
  quantum computers}''.
\newblock \href{https://dx.doi.org/10.1103/PhysRevA.98.032331}{Phys. Rev. A
  {\bf 98}, 032331}~(2018).
\newblock  \href{http://arxiv.org/abs/1803.03326}{arXiv:1803.03326}.

\bibitem{Kaplan:2018vnj}
David~B. Kaplan and Jesse~R. Stryker.
\newblock ``{Gauss\textquoteright{}s law, duality, and the Hamiltonian
  formulation of U(1) lattice gauge theory}''.
\newblock \href{https://dx.doi.org/10.1103/PhysRevD.102.094515}{Phys. Rev. D
  {\bf 102}, 094515}~(2020).
\newblock  \href{http://arxiv.org/abs/1806.08797}{arXiv:1806.08797}.

\bibitem{Kokail:2018eiw}
C.~Kokail, C.~Maier, R.~van Bijnen, T.~Brydges, M.~K. Joshi, P.~Jurcevic, C.~A.
  Muschik, P.~Silvi, R.~Blatt, C.~F. Roos, and P.~Zoller.
\newblock ``{Self-verifying variational quantum simulation of lattice
  models}''.
\newblock \href{https://dx.doi.org/10.1038/s41586-019-1177-4}{Nature {\bf 569},
  355--360}~(2019).
\newblock  \href{http://arxiv.org/abs/1810.03421}{arXiv:1810.03421}.

\bibitem{Stryker:2018efp}
Jesse~R. Stryker.
\newblock ``{Oracles for Gauss's law on digital quantum computers}''.
\newblock \href{https://dx.doi.org/10.1103/PhysRevA.99.042301}{Phys. Rev. A
  {\bf 99}, 042301}~(2019).
\newblock  \href{http://arxiv.org/abs/1812.01617}{arXiv:1812.01617}.

\bibitem{Davoudi:2019bhy}
Zohreh Davoudi, Mohammad Hafezi, Christopher Monroe, Guido Pagano, Alireza
  Seif, and Andrew Shaw.
\newblock ``{Towards analog quantum simulations of lattice gauge theories with
  trapped ions}''.
\newblock \href{https://dx.doi.org/10.1103/PhysRevResearch.2.023015}{Phys. Rev.
  Research {\bf 2}, 023015}~(2020).
\newblock  \href{http://arxiv.org/abs/1908.03210}{arXiv:1908.03210}.

\bibitem{Avkhadiev:2019niu}
A.~Avkhadiev, P.~E. Shanahan, and R.~D. Young.
\newblock ``{Accelerating Lattice Quantum Field Theory Calculations via
  Interpolator Optimization Using Noisy Intermediate-Scale Quantum
  Computing}''.
\newblock \href{https://dx.doi.org/10.1103/PhysRevLett.124.080501}{Phys. Rev.
  Lett. {\bf 124}, 080501}~(2020).
\newblock  \href{http://arxiv.org/abs/1908.04194}{arXiv:1908.04194}.

\bibitem{Magnifico:2019kyj}
Giuseppe Magnifico, Marcello Dalmonte, Paolo Facchi, Saverio Pascazio,
  Francesco~V. Pepe, and Elisa Ercolessi.
\newblock ``{Real Time Dynamics and Confinement in the $\mathbb{Z}_{n}$
  Schwinger-Weyl lattice model for 1+1 QED}''.
\newblock \href{https://dx.doi.org/10.22331/q-2020-06-15-281}{Quantum {\bf 4},
  281}~(2020).
\newblock  \href{http://arxiv.org/abs/1909.04821}{arXiv:1909.04821}.

\bibitem{Luo:2019vmi}
Di~Luo, Jiayu Shen, Michael Highman, Bryan~K. Clark, Brian DeMarco, Aida~X.
  El-Khadra, and Bryce Gadway.
\newblock ``{Framework for simulating gauge theories with dipolar spin
  systems}''.
\newblock \href{https://dx.doi.org/10.1103/PhysRevA.102.032617}{Phys. Rev. A
  {\bf 102}, 032617}~(2020).
\newblock  \href{http://arxiv.org/abs/1912.11488}{arXiv:1912.11488}.

\bibitem{Kharzeev:2020kgc}
Dmitri~E. Kharzeev and Yuta Kikuchi.
\newblock ``Real-time chiral dynamics from a digital quantum simulation''.
\newblock \href{https://dx.doi.org/10.1103/PhysRevResearch.2.023342}{Phys. Rev.
  Research {\bf 2}, 023342}~(2020).
\newblock  \href{http://arxiv.org/abs/2001.00698}{arXiv:2001.00698}.

\bibitem{Shaw2020quantumalgorithms}
Alexander~F. Shaw, Pavel Lougovski, Jesse~R. Stryker, and Nathan Wiebe.
\newblock ``Quantum {A}lgorithms for {S}imulating the {L}attice {S}chwinger
  {M}odel''.
\newblock \href{https://dx.doi.org/10.22331/q-2020-08-10-306}{{Quantum} {\bf
  4}, 306}~(2020).
\newblock  \href{http://arxiv.org/abs/2002.11146}{arXiv:2002.11146}.

\bibitem{PhysRevLett.122.050403}
T.~V. Zache, N.~Mueller, J.~T. Schneider, F.~Jendrzejewski, J.~Berges, and
  P.~Hauke.
\newblock ``{Dynamical Topological Transitions in the Massive Schwinger Model
  with a $\theta$ Term}''.
\newblock \href{https://dx.doi.org/10.1103/PhysRevLett.122.050403}{Phys. Rev.
  Lett. {\bf 122}, 050403}~(2019).
\newblock  \href{http://arxiv.org/abs/1808.07885}{arXiv:1808.07885}.

\bibitem{Yang_2020}
Bing Yang, Hui Sun, Robert Ott, Han-Yi Wang, Torsten~V. Zache, Jad~C. Halimeh,
  Zhen-Sheng Yuan, Philipp Hauke, and Jian-Wei Pan.
\newblock ``Observation of gauge invariance in a 71-site bose–hubbard quantum
  simulator''.
\newblock \href{https://dx.doi.org/10.1038/s41586-020-2910-8}{Nature {\bf 587},
  392–396}~(2020).
\newblock  \href{http://arxiv.org/abs/2003.08945}{arXiv:2003.08945}.

\bibitem{Bender:2020jgr}
Julian Bender, Patrick Emonts, Erez Zohar, and J.~Ignacio Cirac.
\newblock ``{Real-time dynamics in $2+1D$ compact QED using complex periodic
  Gaussian states}''.
\newblock \href{https://dx.doi.org/10.1103/PhysRevResearch.2.043145}{Phys. Rev.
  Research {\bf 2}, 043145}~(2020).
\newblock  \href{http://arxiv.org/abs/2006.10038}{arXiv:2006.10038}.

\bibitem{Haase:2020kaj}
Jan~F. Haase, Luca Dellantonio, Alessio Celi, Danny Paulson, Angus Kan, Karl
  Jansen, and Christine~A. Muschik.
\newblock ``{A resource efficient approach for quantum and classical
  simulations of gauge theories in particle physics}''.
\newblock \href{https://dx.doi.org/10.22331/q-2021-02-04-393}{Quantum {\bf 5},
  393}~(2021).
\newblock  \href{http://arxiv.org/abs/2006.14160}{arXiv:2006.14160}.

\bibitem{Halimeh:2020ecg}
Jad~C. Halimeh, Haifeng Lang, Julius Mildenberger, Zhang Jiang, and Philipp
  Hauke.
\newblock ``{Gauge-Symmetry Protection Using Single-Body Terms}''.
\newblock \href{https://dx.doi.org/10.1103/PRXQuantum.2.040311}{PRX Quantum
  {\bf 2}, 040311}~(2021).
\newblock  \href{http://arxiv.org/abs/2007.00668}{arXiv:2007.00668}.

\bibitem{Robaina:2020aqh}
Daniel Robaina, Mari~Carmen Ba\~nuls, and J.~Ignacio Cirac.
\newblock ``{Simulating $2+1D$ $Z_3$ Lattice Gauge Theory with an Infinite
  Projected Entangled-Pair State}''.
\newblock \href{https://dx.doi.org/10.1103/PhysRevLett.126.050401}{Phys. Rev.
  Lett. {\bf 126}, 050401}~(2021).
\newblock  \href{http://arxiv.org/abs/2007.11630}{arXiv:2007.11630}.

\bibitem{Paulson:2020zjd}
Danny Paulson, Luca Dellantonio, Jan~F. Haase, Alessio Celi, Angus Kan, Andrew
  Jena, Christian Kokail, Rick van Bijnen, Karl Jansen, Peter Zoller, and
  Christine~A. Muschik.
\newblock ``{Towards simulating 2D effects in lattice gauge theories on a
  quantum computer}''.
\newblock \href{https://dx.doi.org/10.1103/PRXQuantum.2.030334}{PRX Quantum
  {\bf 2}, 030334}~(2021).
\newblock  \href{http://arxiv.org/abs/2008.09252}{arXiv:2008.09252}.

\bibitem{VanDamme:2020rur}
Maarten Van~Damme, Jad~C. Halimeh, and Philipp Hauke.
\newblock ``{Gauge-Symmetry Violation Quantum Phase Transition in Lattice Gauge
  Theories}''~(2020).
\newblock  \href{http://arxiv.org/abs/2010.07338}{arXiv:2010.07338}.

\bibitem{Ott:2020ycj}
Robert Ott, Torsten~V. Zache, Fred Jendrzejewski, and J\"urgen Berges.
\newblock ``{Scalable Cold-Atom Quantum Simulator for Two-Dimensional QED}''.
\newblock \href{https://dx.doi.org/10.1103/PhysRevLett.127.130504}{Phys. Rev.
  Lett. {\bf 127}, 130504}~(2021).
\newblock  \href{http://arxiv.org/abs/2012.10432}{arXiv:2012.10432}.

\bibitem{Bauer:2021gup}
Christian~W. Bauer, Marat Freytsis, and Benjamin Nachman.
\newblock ``{Simulating Collider Physics on Quantum Computers Using Effective
  Field Theories}''.
\newblock \href{https://dx.doi.org/10.1103/PhysRevLett.127.212001}{Phys. Rev.
  Lett. {\bf 127}, 212001}~(2021).
\newblock  \href{http://arxiv.org/abs/2102.05044}{arXiv:2102.05044}.

\bibitem{Kan:2021nyu}
Angus Kan, Lena Funcke, Stefan K\"uhn, Luca Dellantonio, Jinglei Zhang, Jan~F.
  Haase, Christine~A. Muschik, and Karl Jansen.
\newblock ``{Investigating a 3+1D Topological $\theta$-Term in the Hamiltonian
  Formulation of Lattice Gauge Theories for Quantum and Classical
  Simulations}''.
\newblock \href{https://dx.doi.org/10.1103/PhysRevD.104.034504}{Phys. Rev. D
  {\bf 104}, 034504}~(2021).
\newblock  \href{http://arxiv.org/abs/2105.06019}{arXiv:2105.06019}.

\bibitem{Aidelsburger:2021mia}
Monika Aidelsburger, Luca Barbiero, Alejandro Bermudez, Titas Chanda, Alexandre
  Dauphin, Daniel González-Cuadra, Przemysław~R. Grzybowski, Simon Hands,
  Fred Jendrzejewski, Johannes Jünemann, Gediminas Juzeliunas, Valentin
  Kasper, Angelo Piga, Shi-Ju Ran, Matteo Rizzi, Gérman Sierra, Luca
  Tagliacozzo, Emanuele Tirrito, Torsten~V. Zache, Jakub Zakrzewski, Erez
  Zohar, and Maciej Lewenstein.
\newblock ``{Cold atoms meet lattice gauge theory}''.
\newblock \href{https://dx.doi.org/10.1098/rsta.2021.0064}{Phil. Trans. Roy.
  Soc. Lond. A {\bf 380}, 20210064}~(2021).
\newblock  \href{http://arxiv.org/abs/2106.03063}{arXiv:2106.03063}.

\bibitem{Meurice:2021pvj}
Yannick Meurice.
\newblock ``{Theoretical methods to design and test quantum simulators for the
  compact Abelian Higgs model}''.
\newblock \href{https://dx.doi.org/10.1103/PhysRevD.104.094513}{Phys. Rev. D
  {\bf 104}, 094513}~(2021).
\newblock  \href{http://arxiv.org/abs/2107.11366}{arXiv:2107.11366}.

\bibitem{Mueller:2021gxd}
Niklas Mueller, Torsten~V. Zache, and Robert Ott.
\newblock ``{Thermalization of Gauge Theories from their Entanglement
  Spectrum}''.
\newblock \href{https://dx.doi.org/10.1103/PhysRevLett.129.011601}{Phys. Rev.
  Lett. {\bf 129}, 011601}~(2022).
\newblock  \href{http://arxiv.org/abs/2107.11416}{arXiv:2107.11416}.

\bibitem{Riechert:2021ink}
Hannes Riechert, Jad~C. Halimeh, Valentin Kasper, Landry Bretheau, Erez Zohar,
  Philipp Hauke, and Fred Jendrzejewski.
\newblock ``{Engineering a U(1) lattice gauge theory in classical electric
  circuits}''.
\newblock \href{https://dx.doi.org/10.1103/PhysRevB.105.205141}{Phys. Rev. B
  {\bf 105}, 205141}~(2022).
\newblock  \href{http://arxiv.org/abs/2108.01086}{arXiv:2108.01086}.

\bibitem{Halimeh:2021lnv}
Jad~C. Halimeh, Lukas Homeier, Christian Schweizer, Monika Aidelsburger,
  Philipp Hauke, and Fabian Grusdt.
\newblock ``{Stabilizing Lattice Gauge Theories Through Simplified Local
  Pseudogenerators}''.
\newblock \href{https://dx.doi.org/10.1103/PhysRevResearch.4.033120}{Phys. Rev.
  Research {\bf 4}, 033120}~(2022).
\newblock  \href{http://arxiv.org/abs/2108.02203}{arXiv:2108.02203}.

\bibitem{Zhang:2021bjq}
Jinglei Zhang, Ryan Ferguson, Stefan K\"uhn, Jan~F. Haase, C.~M. Wilson, Karl
  Jansen, and Christine~A. Muschik.
\newblock ``{Simulating gauge theories with variational quantum eigensolvers in
  superconducting microwave cavities}''~(2021).
\newblock  \href{http://arxiv.org/abs/2108.08248}{arXiv:2108.08248}.

\bibitem{Gustafson:2021jtq}
Erik Gustafson, Burt Holzman, James Kowalkowski, Henry Lamm, Andy C.~Y. Li,
  Gabriel Perdue, Sergio Boixo, Sergei Isakov, Orion Martin, Ross Thomson,
  Catherine~Vollgraff Heidweiller, Jackson Beall, Martin Ganahl, Guifre Vidal,
  and Evan Peters.
\newblock ``{Large scale multi-node simulations of $\mathbb{Z}_2$ gauge theory
  quantum circuits using Google Cloud Platform}''.
\newblock In {IEEE/ACM Second International Workshop on Quantum Computing
  Software}.
\newblock ~(2021).
\newblock  \href{http://arxiv.org/abs/2110.07482}{arXiv:2110.07482}.

\bibitem{Thompson:2021eze}
Shane Thompson and George Siopsis.
\newblock ``{Quantum computation of phase transition in the massive Schwinger
  model}''.
\newblock \href{https://dx.doi.org/10.1088/2058-9565/ac5f5a}{Quantum Sci.
  Technol. {\bf 7}, 035001}~(2022).
\newblock  \href{http://arxiv.org/abs/2110.13046}{arXiv:2110.13046}.

\bibitem{Kan:2021blb}
Angus Kan, Lena Funcke, Stefan K\"uhn, Luca Dellantonio, Jinglei Zhang, Jan~F.
  Haase, Christine~A. Muschik, and Karl Jansen.
\newblock ``{3+1D $\theta$-Term on the Lattice from the Hamiltonian
  Perspective}''.
\newblock \href{https://dx.doi.org/10.22323/1.396.0112}{PoS {\bf LATTICE2021},
  112}~(2022).
\newblock  \href{http://arxiv.org/abs/2111.02238}{arXiv:2111.02238}.

\bibitem{Ashkenazi:2021ieg}
Shachar Ashkenazi and Erez Zohar.
\newblock ``{Duality as a feasible physical transformation for quantum
  simulation}''.
\newblock \href{https://dx.doi.org/10.1103/PhysRevA.105.022431}{Phys. Rev. A
  {\bf 105}, 022431}~(2022).
\newblock  \href{http://arxiv.org/abs/2111.04765}{arXiv:2111.04765}.

\bibitem{Bauer:2021gek}
Christian~W. Bauer and Dorota~M. Grabowska.
\newblock ``{Efficient Representation for Simulating U(1) Gauge Theories on
  Digital Quantum Computers at All Values of the Coupling}''~(2021).
\newblock  \href{http://arxiv.org/abs/2111.08015}{arXiv:2111.08015}.

\bibitem{Nguyen:2021hyk}
Nhung~H. Nguyen, Minh~C. Tran, Yingyue Zhu, Alaina~M. Green, C.~Huerta
  Alderete, Zohreh Davoudi, and Norbert~M. Linke.
\newblock ``{Digital Quantum Simulation of the Schwinger Model and Symmetry
  Protection with Trapped Ions}''.
\newblock \href{https://dx.doi.org/10.1103/PRXQuantum.3.020324}{PRX Quantum
  {\bf 3}, 020324}~(2022).
\newblock  \href{http://arxiv.org/abs/2112.14262}{arXiv:2112.14262}.

\bibitem{Halimeh:2022rwu}
Jad~C. Halimeh, Luca Barbiero, Philipp Hauke, Fabian Grusdt, and Annabelle
  Bohrdt.
\newblock ``{Robust quantum many-body scars in lattice gauge
  theories}''~(2022).
\newblock  \href{http://arxiv.org/abs/2203.08828}{arXiv:2203.08828}.

\bibitem{Mildenberger:2022jqr}
Julius Mildenberger, Wojciech Mruczkiewicz, Jad~C. Halimeh, Zhang Jiang, and
  Philipp Hauke.
\newblock ``{Probing confinement in a $\mathbb{Z}_2$ lattice gauge theory on a
  quantum computer}''~(2022).
\newblock  \href{http://arxiv.org/abs/2203.08905}{arXiv:2203.08905}.

\bibitem{Halimeh:2022pkw}
Jad~C. Halimeh, Ian~P. McCulloch, Bing Yang, and Philipp Hauke.
\newblock ``{Tuning the Topological $\theta$-Angle in Cold-Atom Quantum
  Simulators of Gauge Theories}''~(2022).
\newblock  \href{http://arxiv.org/abs/2204.06570}{arXiv:2204.06570}.

\bibitem{Greenberg:2022kzy}
Tomer Greenberg, Guy Pardo, Aryeh Fortinsky, and Erez Zohar.
\newblock ``{Resource-Efficient Quantum Simulation of Lattice Gauge Theories in
  Arbitrary Dimensions: Solving for Gauss' Law and Fermion
  Elimination}''~(2022).
\newblock  \href{http://arxiv.org/abs/2206.00685}{arXiv:2206.00685}.

\bibitem{Grabowska:2022uos}
Dorota~M. Grabowska, Christopher Kane, Benjamin Nachman, and Christian~W.
  Bauer.
\newblock ``{Overcoming exponential scaling with system size in Trotter-Suzuki
  implementations of constrained Hamiltonians: 2+1 U(1) lattice gauge
  theories}''~(2022).
\newblock  \href{http://arxiv.org/abs/2208.03333}{arXiv:2208.03333}.

\bibitem{Davoudi:2022uzo}
Zohreh Davoudi, Niklas Mueller, and Connor Powers.
\newblock ``{Toward Quantum Computing Phase Diagrams of Gauge Theories with
  Thermal Pure Quantum States}''~(2022).
\newblock  \href{http://arxiv.org/abs/2208.13112}{arXiv:2208.13112}.

\bibitem{https://doi.org/10.48550/arxiv.2206.12454}
Giuseppe Clemente, Arianna Crippa, and Karl Jansen.
\newblock ``Strategies for the determination of the running coupling of
  $(2+1)$-dimensional qed with quantum computing''~(2022).

\bibitem{Brower:1997ha}
R.~Brower, S.~Chandrasekharan, and U.~J. Wiese.
\newblock ``{QCD as a quantum link model}''.
\newblock \href{https://dx.doi.org/10.1103/PhysRevD.60.094502}{Phys. Rev. D
  {\bf 60}, 094502}~(1999).
\newblock  \href{http://arxiv.org/abs/hep-th/9704106}{arXiv:hep-th/9704106}.

\bibitem{PhysRevA.73.022328}
Tim Byrnes and Yoshihisa Yamamoto.
\newblock ``Simulating lattice gauge theories on a quantum computer''.
\newblock \href{https://dx.doi.org/10.1103/PhysRevA.73.022328}{Phys. Rev. A
  {\bf 73}, 022328}~(2006).
\newblock
  \href{http://arxiv.org/abs/quant-ph/0510027}{arXiv:quant-ph/0510027}.

\bibitem{Banerjee:2012xg}
D.~Banerjee, M.~B\"ogli, M.~Dalmonte, E.~Rico, P.~Stebler, U.~J. Wiese, and
  P.~Zoller.
\newblock ``{Atomic Quantum Simulation of U(N) and SU(N) Non-Abelian Lattice
  Gauge Theories}''.
\newblock \href{https://dx.doi.org/10.1103/PhysRevLett.110.125303}{Phys. Rev.
  Lett. {\bf 110}, 125303}~(2013).
\newblock  \href{http://arxiv.org/abs/1211.2242}{arXiv:1211.2242}.

\bibitem{Tagliacozzo:2012df}
L.~Tagliacozzo, A.~Celi, P.~Orland, and M.~Lewenstein.
\newblock ``{Simulations of non-Abelian gauge theories with optical
  lattices}''.
\newblock \href{https://dx.doi.org/10.1038/ncomms3615}{Nat Commun {\bf 4},
  2615}~(2013).
\newblock  \href{http://arxiv.org/abs/1211.2704}{arXiv:1211.2704}.

\bibitem{Zohar:2012xf}
Erez Zohar, J.~Ignacio Cirac, and Benni Reznik.
\newblock ``{Cold-Atom Quantum Simulator for SU(2) Yang-Mills Lattice Gauge
  Theory}''.
\newblock \href{https://dx.doi.org/10.1103/PhysRevLett.110.125304}{Phys. Rev.
  Lett. {\bf 110}, 125304}~(2013).
\newblock  \href{http://arxiv.org/abs/1211.2241}{arXiv:1211.2241}.

\bibitem{Wiese:2013uua}
Uwe-Jens Wiese.
\newblock ``{Ultracold Quantum Gases and Lattice Systems: Quantum Simulation of
  Lattice Gauge Theories}''.
\newblock \href{https://dx.doi.org/10.1002/andp.201300104}{Annalen Phys. {\bf
  525}, 777--796}~(2013).
\newblock  \href{http://arxiv.org/abs/1305.1602}{arXiv:1305.1602}.

\bibitem{Zohar:2016iic}
Erez Zohar, Alessandro Farace, Benni Reznik, and J.~Ignacio Cirac.
\newblock ``{Digital lattice gauge theories}''.
\newblock \href{https://dx.doi.org/10.1103/PhysRevA.95.023604}{Phys. Rev. A
  {\bf 95}, 023604}~(2017).
\newblock  \href{http://arxiv.org/abs/1607.08121}{arXiv:1607.08121}.

\bibitem{Banuls:2017ena}
Mari~Carmen Ba\~nuls, Krzysztof Cichy, J.~Ignacio Cirac, Karl Jansen, and
  Stefan K\"uhn.
\newblock ``{Efficient basis formulation for 1+1 dimensional SU(2) lattice
  gauge theory: Spectral calculations with matrix product states}''.
\newblock \href{https://dx.doi.org/10.1103/PhysRevX.7.041046}{Phys. Rev. X {\bf
  7}, 041046}~(2017).
\newblock  \href{http://arxiv.org/abs/1707.06434}{arXiv:1707.06434}.

\bibitem{Alexandru:2019nsa}
Andrei Alexandru, Paulo~F. Bedaque, Siddhartha Harmalkar, Henry Lamm, Scott
  Lawrence, and Neill~C. Warrington.
\newblock ``{Gluon Field Digitization for Quantum Computers}''.
\newblock \href{https://dx.doi.org/10.1103/PhysRevD.100.114501}{Phys. Rev. D
  {\bf 100}, 114501}~(2019).
\newblock  \href{http://arxiv.org/abs/1906.11213}{arXiv:1906.11213}.

\bibitem{Klco:2019evd}
Natalie Klco, Jesse~R. Stryker, and Martin~J. Savage.
\newblock ``{SU(2) non-Abelian gauge field theory in one dimension on digital
  quantum computers}''.
\newblock \href{https://dx.doi.org/10.1103/PhysRevD.101.074512}{Phys. Rev. D
  {\bf 101}, 074512}~(2020).
\newblock  \href{http://arxiv.org/abs/1908.06935}{arXiv:1908.06935}.

\bibitem{Banuls:2019bmf}
Mari~Carmen Ba{\~{n}}uls, Rainer Blatt, Jacopo Catani, Alessio Celi,
  Juan~Ignacio Cirac, Marcello Dalmonte, Leonardo Fallani, Karl Jansen, Maciej
  Lewenstein, Simone Montangero, Christine~A. Muschik, Benni Reznik, Enrique
  Rico, Luca Tagliacozzo, Karel~Van Acoleyen, Frank Verstraete, Uwe-Jens Wiese,
  Matthew Wingate, Jakub Zakrzewski, and Peter Zoller.
\newblock ``{Simulating Lattice Gauge Theories within Quantum Technologies}''.
\newblock \href{https://dx.doi.org/10.1140/epjd/e2020-100571-8}{Eur. Phys. J. D
  {\bf 74}, 165}~(2020).
\newblock  \href{http://arxiv.org/abs/1911.00003}{arXiv:1911.00003}.

\bibitem{Ji:2020kjk}
Yao Ji, Henry Lamm, and Shuchen Zhu.
\newblock ``{Gluon Field Digitization via Group Space Decimation for Quantum
  Computers}''.
\newblock \href{https://dx.doi.org/10.1103/PhysRevD.102.114513}{Phys. Rev. D
  {\bf 102}, 114513}~(2020).
\newblock  \href{http://arxiv.org/abs/2005.14221}{arXiv:2005.14221}.

\bibitem{Halimeh:2020djb}
Jad~C. Halimeh, Valentin Kasper, and Philipp Hauke.
\newblock ``{Fate of Lattice Gauge Theories Under Decoherence}''~(2020).
\newblock  \href{http://arxiv.org/abs/2009.07848}{arXiv:2009.07848}.

\bibitem{Kasper:2020owz}
Valentin Kasper, Torsten~V. Zache, Fred Jendrzejewski, Maciej Lewenstein, and
  Erez Zohar.
\newblock ``{Non-Abelian gauge invariance from dynamical decoupling}''~(2020).
\newblock  \href{http://arxiv.org/abs/2012.08620}{arXiv:2012.08620}.

\bibitem{Ciavarella:2021nmj}
Anthony Ciavarella, Natalie Klco, and Martin~J. Savage.
\newblock ``{Trailhead for quantum simulation of SU(3) Yang-Mills lattice gauge
  theory in the local multiplet basis}''.
\newblock \href{https://dx.doi.org/10.1103/PhysRevD.103.094501}{Phys. Rev. D
  {\bf 103}, 094501}~(2021).
\newblock  \href{http://arxiv.org/abs/2101.10227}{arXiv:2101.10227}.

\bibitem{ARahman:2021ktn}
Sarmed A~Rahman, Randy Lewis, Emanuele Mendicelli, and Sarah Powell.
\newblock ``{SU(2) lattice gauge theory on a quantum annealer}''.
\newblock \href{https://dx.doi.org/10.1103/PhysRevD.104.034501}{Phys. Rev. D
  {\bf 104}, 034501}~(2021).
\newblock  \href{http://arxiv.org/abs/2103.08661}{arXiv:2103.08661}.

\bibitem{Atas:2021ext}
Yasar~Y. Atas, Jinglei Zhang, Randy Lewis, Amin Jahanpour, Jan~F. Haase, and
  Christine~A. Muschik.
\newblock ``{SU(2) hadrons on a quantum computer via a variational approach}''.
\newblock \href{https://dx.doi.org/10.1038/s41467-021-26825-4}{Nat Commun {\bf
  12}, 6499}~(2021).
\newblock  \href{http://arxiv.org/abs/2102.08920}{arXiv:2102.08920}.

\bibitem{Davoudi:2021ney}
Zohreh Davoudi, Norbert~M. Linke, and Guido Pagano.
\newblock ``{Toward simulating quantum field theories with controlled
  phonon-ion dynamics: A hybrid analog-digital approach}''.
\newblock \href{https://dx.doi.org/10.1103/PhysRevResearch.3.043072}{Phys. Rev.
  Research {\bf 3}, 043072}~(2021).
\newblock  \href{http://arxiv.org/abs/2104.09346}{arXiv:2104.09346}.

\bibitem{Stryker:2021asy}
Jesse~R. Stryker.
\newblock ``{Shearing approach to gauge invariant Trotterization}''~(2021).
\newblock  \href{http://arxiv.org/abs/2105.11548}{arXiv:2105.11548}.

\bibitem{Zohar:2021nyc}
Erez Zohar.
\newblock ``{Quantum simulation of lattice gauge theories in more than one
  space dimension\textemdash{}requirements, challenges and methods}''.
\newblock \href{https://dx.doi.org/10.1098/rsta.2021.0069}{Phil. Trans. A.
  Math. Phys. Eng. Sci. {\bf 380}, 20210069}~(2021).
\newblock  \href{http://arxiv.org/abs/2106.04609}{arXiv:2106.04609}.

\bibitem{Halimeh:2021vzf}
Jad~C. Halimeh, Haifeng Lang, and Philipp Hauke.
\newblock ``{Gauge protection in non-abelian lattice gauge theories}''.
\newblock \href{https://dx.doi.org/10.1088/1367-2630/ac5564}{New J. Phys. {\bf
  24}, 033015}~(2022).
\newblock  \href{http://arxiv.org/abs/2106.09032}{arXiv:2106.09032}.

\bibitem{Wiese:2021djl}
Uwe-Jens Wiese.
\newblock ``{From quantum link models to D-theory: a resource efficient
  framework for the quantum simulation and computation of gauge theories}''.
\newblock \href{https://dx.doi.org/10.1098/rsta.2021.0068}{Phil. Trans. A.
  Math. Phys. Eng. Sci. {\bf 380}, 20210068}~(2021).
\newblock  \href{http://arxiv.org/abs/2107.09335}{arXiv:2107.09335}.

\bibitem{Alam:2021uuq}
M.~Sohaib Alam, Stuart Hadfield, Henry Lamm, and Andy C.~Y. Li.
\newblock ``{Primitive quantum gates for dihedral gauge theories}''.
\newblock \href{https://dx.doi.org/10.1103/PhysRevD.105.114501}{Phys. Rev. D
  {\bf 105}, 114501}~(2022).
\newblock  \href{http://arxiv.org/abs/2108.13305}{arXiv:2108.13305}.

\bibitem{Funcke:2021aps}
Lena Funcke, Tobias Hartung, Karl Jansen, Stefan K\"uhn, Manuel Schneider,
  Paolo Stornati, and Xiaoyang Wang.
\newblock ``{Towards quantum simulations in particle physics and beyond on
  noisy intermediate-scale quantum devices}''.
\newblock \href{https://dx.doi.org/10.1098/rsta.2021.0062}{Phil. Trans. A.
  Math. Phys. Eng. Sci. {\bf 380}, 20210062}~(2021).
\newblock  \href{http://arxiv.org/abs/2110.03809}{arXiv:2110.03809}.

\bibitem{VanDamme:2021njp}
Maarten Van~Damme, Julius Mildenberger, Fabian Grusdt, Philipp Hauke, and
  Jad~C. Halimeh.
\newblock ``{Suppressing nonperturbative gauge errors in the thermodynamic
  limit using local pseudogenerators}''~(2021).
\newblock  \href{http://arxiv.org/abs/2110.08041}{arXiv:2110.08041}.

\bibitem{Alexandrou:2021ynh}
Constantia Alexandrou, Lena Funcke, Tobias Hartung, Karl Jansen, Stefan K\"uhn,
  Georgios Polykratis, Paolo Stornati, Xiaoyang Wang, and Tom Weber.
\newblock ``{Investigating the variance increase of readout error mitigation
  through classical bit-flip correction on IBM and Rigetti quantum
  computers}''.
\newblock \href{https://dx.doi.org/10.22323/1.396.0243}{PoS {\bf LATTICE2021},
  243}~(2022).
\newblock  \href{http://arxiv.org/abs/2111.05026}{arXiv:2111.05026}.

\bibitem{Ciavarella:2021lel}
Anthony~N. Ciavarella and Ivan~A. Chernyshev.
\newblock ``{Preparation of the SU(3) lattice Yang-Mills vacuum with
  variational quantum methods}''.
\newblock \href{https://dx.doi.org/10.1103/PhysRevD.105.074504}{Phys. Rev. D
  {\bf 105}, 074504}~(2022).
\newblock  \href{http://arxiv.org/abs/2112.09083}{arXiv:2112.09083}.

\bibitem{Hartung:2022hoz}
Tobias Hartung, Timo Jakobs, Karl Jansen, Johann Ostmeyer, and Carsten Urbach.
\newblock ``{Digitising SU(2) gauge fields and the freezing transition}''.
\newblock \href{https://dx.doi.org/10.1140/epjc/s10052-022-10192-5}{Eur. Phys.
  J. C {\bf 82}, 237}~(2022).
\newblock  \href{http://arxiv.org/abs/2201.09625}{arXiv:2201.09625}.

\bibitem{Illa:2022jqb}
Marc Illa and Martin~J. Savage.
\newblock ``{Basic Elements for Simulations of Standard Model Physics with
  Quantum Annealers: Multigrid and Clock States}''.
\newblock \href{https://dx.doi.org/10.1103/PhysRevA.106.052605}{Phys. Rev. A
  {\bf 106}, 052605}~(2022).
\newblock  \href{http://arxiv.org/abs/2202.12340}{arXiv:2202.12340}.

\bibitem{Ji:2022qvr}
Yao Ji, Henry Lamm, and Shuchen Zhu.
\newblock ``{Gluon Digitization via Character Expansion for Quantum
  Computers}''~(2022).
\newblock  \href{http://arxiv.org/abs/2203.02330}{arXiv:2203.02330}.

\bibitem{Carena:2022kpg}
Marcela Carena, Henry Lamm, Ying-Ying Li, and Wanqiang Liu.
\newblock ``{Improved Hamiltonians for Quantum Simulations of Gauge
  Theories}''.
\newblock \href{https://dx.doi.org/10.1103/PhysRevLett.129.051601}{Phys. Rev.
  Lett. {\bf 129}, 051601}~(2022).
\newblock  \href{http://arxiv.org/abs/2203.02823}{arXiv:2203.02823}.

\bibitem{Ciavarella:2022zhe}
Anthony Ciavarella, Natalie Klco, and Martin~J. Savage.
\newblock ``{Some Conceptual Aspects of Operator Design for Quantum Simulations
  of Non-Abelian Lattice Gauge Theories}''.
\newblock In Proceedings of the 2021 Quantum Simulation for Strong Interactions
  (QuaSi) Workshops.
\newblock ~(2022).
\newblock  \href{http://arxiv.org/abs/2203.11988}{arXiv:2203.11988}.

\bibitem{Bauer:2022hpo}
Christian~W. Bauer, Zohreh Davoudi, A.~Baha Balantekin, Tanmoy Bhattacharya,
  Marcela Carena, Wibe~A. de~Jong, Patrick Draper, Aida El-Khadra, Nate
  Gemelke, Masanori Hanada, Dmitri Kharzeev, Henry Lamm, Ying-Ying Li, Junyu
  Liu, Mikhail Lukin, Yannick Meurice, Christopher Monroe, Benjamin Nachman,
  Guido Pagano, John Preskill, Enrico Rinaldi, Alessandro Roggero, David~I.
  Santiago, Martin~J. Savage, Irfan Siddiqi, George Siopsis, David Van~Zanten,
  Nathan Wiebe, Yukari Yamauchi, Kübra Yeter-Aydeniz, and Silvia Zorzetti.
\newblock ``{Quantum Simulation for High Energy Physics}''~(2022).
\newblock  \href{http://arxiv.org/abs/2204.03381}{arXiv:2204.03381}.

\bibitem{Raychowdhury:2022wbi}
Indrakshi Raychowdhury, Zohreh Davoudi, and Andrew Shaw.
\newblock ``{Exploring different Formulations of non-Abelian Lattice Gauge
  Theories for Hamiltonian simulation}''.
\newblock \href{https://dx.doi.org/10.22323/1.396.0277}{PoS {\bf LATTICE2021},
  277}~(2022).

\bibitem{Rahman:2022rlg}
Sarmed A~Rahman, Randy Lewis, Emanuele Mendicelli, and Sarah Powell.
\newblock ``{Self-mitigating Trotter circuits for SU(2) lattice gauge theory on
  a quantum computer}''~(2022).
\newblock  \href{http://arxiv.org/abs/2205.09247}{arXiv:2205.09247}.

\bibitem{Farrell:2022wyt}
Roland~C. Farrell, Ivan~A. Chernyshev, Sarah J.~M. Powell, Nikita~A.
  Zemlevskiy, Marc Illa, and Martin~J. Savage.
\newblock ``{Preparations for Quantum Simulations of Quantum Chromodynamics in
  1+1 Dimensions: (I) Axial Gauge}''~(2022).
\newblock  \href{http://arxiv.org/abs/2207.01731}{arXiv:2207.01731}.

\bibitem{Atas:2022dqm}
Yasar~Y. Atas, Jan~F. Haase, Jinglei Zhang, Victor Wei, Sieglinde M.~L.
  Pfaendler, Randy Lewis, and Christine~A. Muschik.
\newblock ``{Real-time evolution of SU(3) hadrons on a quantum
  computer}''~(2022).
\newblock  \href{http://arxiv.org/abs/2207.03473}{arXiv:2207.03473}.

\bibitem{Carena:2022hpz}
Marcela Carena, Erik~J. Gustafson, Henry Lamm, Ying-Ying Li, and Wanqiang Liu.
\newblock ``{Gauge Theory Couplings on Anisotropic Lattices}''~(2022).
\newblock  \href{http://arxiv.org/abs/2208.10417}{arXiv:2208.10417}.

\bibitem{Gustafson:2022xdt}
Erik~J. Gustafson, Henry Lamm, Felicity Lovelace, and Damian Musk.
\newblock ``{Primitive Quantum Gates for an SU(2) Discrete Subgroup:
  BT}''~(2022).
\newblock  \href{http://arxiv.org/abs/2208.12309}{arXiv:2208.12309}.

\bibitem{Avkhadiev:2022ttx}
A.~Avkhadiev, P.~E. Shanahan, and R.~D. Young.
\newblock ``{Strategies for quantum-optimized construction of interpolating
  operators in classical simulations of lattice quantum field
  theories}''~(2022).
\newblock  \href{http://arxiv.org/abs/2209.01209}{arXiv:2209.01209}.

\bibitem{Farrell:2022vyh}
Roland~C. Farrell, Ivan~A. Chernyshev, Sarah J.~M. Powell, Nikita~A.
  Zemlevskiy, Marc Illa, and Martin~J. Savage.
\newblock ``{Preparations for Quantum Simulations of Quantum Chromodynamics in
  1+1 Dimensions: (II) Single-Baryon Beta-Decay in Real Time}''~(2022).
\newblock  \href{http://arxiv.org/abs/2209.10781}{arXiv:2209.10781}.

\bibitem{Mishra:2019xbh}
Chinmay Mishra, Shane Thompson, Raphael Pooser, and George Siopsis.
\newblock ``{Quantum computation of an interacting fermionic model}''.
\newblock \href{https://dx.doi.org/10.1088/2058-9565/ab8f63}{Quantum Sci.
  Technol. {\bf 5}, 035010}~(2020).
\newblock  \href{http://arxiv.org/abs/1912.07767}{arXiv:1912.07767}.

\bibitem{Perlin:2021xux}
Michael~A. Perlin, Diego Barberena, Mikhail Mamaev, Bhuvanesh Sundar, Robert~J.
  Lewis-Swan, and Ana~Maria Rey.
\newblock ``{Engineering infinite-range SU(n) interactions with
  spin-orbit-coupled fermions in an optical lattice}''.
\newblock \href{https://dx.doi.org/10.1103/PhysRevA.105.023326}{Phys. Rev. A
  {\bf 105}, 023326}~(2022).
\newblock  \href{http://arxiv.org/abs/2109.11019}{arXiv:2109.11019}.

\bibitem{Bringewatt:2022zgq}
Jacob Bringewatt and Zohreh Davoudi.
\newblock ``{Parallelization techniques for quantum simulation of fermionic
  systems}''~(2022).
\newblock  \href{http://arxiv.org/abs/2207.12470}{arXiv:2207.12470}.

\bibitem{Asaduzzaman:2022bpi}
Muhammad Asaduzzaman, Simon Catterall, Goksu~Can Toga, Yannick Meurice, and Ryo
  Sakai.
\newblock ``{Quantum Simulation of the N flavor Gross-Neveu Model}''~(2022).
\newblock  \href{http://arxiv.org/abs/2208.05906}{arXiv:2208.05906}.

\bibitem{Yeter-Aydeniz:2018mix}
Kubra Yeter-Aydeniz, Eugene~F. Dumitrescu, Alex~J. McCaskey, Ryan~S. Bennink,
  Raphael~C. Pooser, and George Siopsis.
\newblock ``{Scalar Quantum Field Theories as a Benchmark for Near-Term Quantum
  Computers}''.
\newblock \href{https://dx.doi.org/10.1103/PhysRevA.99.032306}{Phys. Rev. A
  {\bf 99}, 032306}~(2019).
\newblock  \href{http://arxiv.org/abs/1811.12332}{arXiv:1811.12332}.

\bibitem{Klco:2019xro}
Natalie Klco and Martin~J. Savage.
\newblock ``{Minimally entangled state preparation of localized wave functions
  on quantum computers}''.
\newblock \href{https://dx.doi.org/10.1103/PhysRevA.102.012612}{Phys. Rev. A
  {\bf 102}, 012612}~(2020).
\newblock  \href{http://arxiv.org/abs/1904.10440}{arXiv:1904.10440}.

\bibitem{Klco:2019yrb}
Natalie Klco and Martin~J. Savage.
\newblock ``{Systematically Localizable Operators for Quantum Simulations of
  Quantum Field Theories}''.
\newblock \href{https://dx.doi.org/10.1103/PhysRevA.102.012619}{Phys. Rev. A
  {\bf 102}, 012619}~(2020).
\newblock  \href{http://arxiv.org/abs/1912.03577}{arXiv:1912.03577}.

\bibitem{Barata:2020jtq}
Jo\~ao Barata, Niklas Mueller, Andrey Tarasov, and Raju Venugopalan.
\newblock ``{Single-particle digitization strategy for quantum computation of a
  $\phi^4$ scalar field theory}''.
\newblock \href{https://dx.doi.org/10.1103/PhysRevA.103.042410}{Phys. Rev. A
  {\bf 103}, 042410}~(2021).
\newblock  \href{http://arxiv.org/abs/2012.00020}{arXiv:2012.00020}.

\bibitem{Yeter-Aydeniz:2021mol}
K\"ubra Yeter-Aydeniz, Eleftherios Moschandreou, and George Siopsis.
\newblock ``{Quantum imaginary-time evolution algorithm for quantum field
  theories with continuous variables}''.
\newblock \href{https://dx.doi.org/10.1103/PhysRevA.105.012412}{Phys. Rev. A
  {\bf 105}, 012412}~(2022).
\newblock  \href{http://arxiv.org/abs/2107.00791}{arXiv:2107.00791}.

\bibitem{Deliyannis:2021che}
Plato Deliyannis, Marat Freytsis, Benjamin Nachman, and Christian~W. Bauer.
\newblock ``{Practical considerations for the preparation of multivariate
  Gaussian states on quantum computers}''~(2021).
\newblock  \href{http://arxiv.org/abs/2109.10918}{arXiv:2109.10918}.

\bibitem{Caspar:2022llo}
Stephan Caspar and Hersh Singh.
\newblock ``{From Asymptotic Freedom to \ensuremath{\theta} Vacua: Qubit
  Embeddings of the O(3) Nonlinear \ensuremath{\sigma} Model}''.
\newblock \href{https://dx.doi.org/10.1103/PhysRevLett.129.022003}{Phys. Rev.
  Lett. {\bf 129}, 022003}~(2022).
\newblock  \href{http://arxiv.org/abs/2203.15766}{arXiv:2203.15766}.

\bibitem{Dumitrescu:2018njn}
E.~F. Dumitrescu, A.~J. McCaskey, G.~Hagen, G.~R. Jansen, T.~D. Morris,
  T.~Papenbrock, R.~C. Pooser, D.~J. Dean, and P.~Lougovski.
\newblock ``{Cloud Quantum Computing of an Atomic Nucleus}''.
\newblock \href{https://dx.doi.org/10.1103/PhysRevLett.120.210501}{Phys. Rev.
  Lett. {\bf 120}, 210501}~(2018).
\newblock  \href{http://arxiv.org/abs/1801.03897}{arXiv:1801.03897}.

\bibitem{Lu:2018pjk}
Hsuan-Hao Lu, Natalie Klco, Joseph~M. Lukens, Titus~D. Morris, Aaina Bansal,
  Andreas Ekstr\"om, Gaute Hagen, Thomas Papenbrock, Andrew~M. Weiner,
  Martin~J. Savage, and Pavel Lougovski.
\newblock ``Simulations of subatomic many-body physics on a quantum frequency
  processor''.
\newblock \href{https://dx.doi.org/10.1103/PhysRevA.100.012320}{Phys. Rev. A
  {\bf 100}, 012320}~(2019).
\newblock  \href{http://arxiv.org/abs/1810.03959}{arXiv:1810.03959}.

\bibitem{Cloet:2019wre}
John Arrington et~al.
\newblock ``{Opportunities for Nuclear Physics \& Quantum Information
  Science}''.
\newblock In Ian~C. Clo\"et and Matthew~R. Dietrich, editors, {Intersections
  between Nuclear Physics and Quantum Information}.
\newblock ~(2019).
\newblock  \href{http://arxiv.org/abs/1903.05453}{arXiv:1903.05453}.

\bibitem{Shehab:2019gfn}
Omar Shehab, Kevin~A. Landsman, Yunseong Nam, Daiwei Zhu, Norbert~M. Linke,
  Matthew~J. Keesan, Raphael~C. Pooser, and Christopher~R. Monroe.
\newblock ``{Toward convergence of effective field theory simulations on
  digital quantum computers}''.
\newblock \href{https://dx.doi.org/10.1103/PhysRevA.100.062319}{Phys. Rev. A
  {\bf 100}, 062319}~(2019).
\newblock  \href{http://arxiv.org/abs/1904.04338}{arXiv:1904.04338}.

\bibitem{Mueller:2020vha}
Niklas Mueller, Andrey Tarasov, and Raju Venugopalan.
\newblock ``{Computing real time correlation functions on a hybrid
  classical/quantum computer}''.
\newblock \href{https://dx.doi.org/10.1016/j.nuclphysa.2020.121889}{Nucl. Phys.
  A {\bf 1005}, 121889}~(2021).
\newblock  \href{http://arxiv.org/abs/2001.11145}{arXiv:2001.11145}.

\bibitem{Knaute:2021xna}
Johannes Knaute and Philipp Hauke.
\newblock ``{Relativistic meson spectra on ion-trap quantum simulators}''.
\newblock \href{https://dx.doi.org/10.1103/PhysRevA.105.022616}{Phys. Rev. A
  {\bf 105}, 022616}~(2022).
\newblock  \href{http://arxiv.org/abs/2107.09071}{arXiv:2107.09071}.

\bibitem{Yeter-Aydeniz:2021olz}
K\"ubra Yeter-Aydeniz, Shikha Bangar, George Siopsis, and Raphael~C. Pooser.
\newblock ``{Collective neutrino oscillations on a quantum computer}''.
\newblock \href{https://dx.doi.org/10.1007/s11128-021-03348-x}{Quant. Inf.
  Proc. {\bf 21}, 84}~(2022).
\newblock  \href{http://arxiv.org/abs/2104.03273}{arXiv:2104.03273}.

\bibitem{Gustafson:2019mpk}
Erik Gustafson, Yannick Meurice, and Judah Unmuth-Yockey.
\newblock ``{Quantum simulation of scattering in the quantum Ising model}''.
\newblock \href{https://dx.doi.org/10.1103/PhysRevD.99.094503}{Phys. Rev. D
  {\bf 99}, 094503}~(2019).
\newblock  \href{http://arxiv.org/abs/1901.05944}{arXiv:1901.05944}.

\bibitem{Bauer:2019qxa}
Christian~W. Bauer, Wibe~A. de~Jong, Benjamin Nachman, and Davide Provasoli.
\newblock ``{Quantum Algorithm for High Energy Physics Simulations}''.
\newblock \href{https://dx.doi.org/10.1103/PhysRevLett.126.062001}{Phys. Rev.
  Lett. {\bf 126}, 062001}~(2021).
\newblock  \href{http://arxiv.org/abs/1904.03196}{arXiv:1904.03196}.

\bibitem{Gustafson:2021mky}
Erik Gustafson, Patrick Dreher, Zheyue Hang, and Yannick Meurice.
\newblock ``{Indexed improvements for real-time trotter evolution of a (1 + 1)
  field theory using NISQ quantum computers}''.
\newblock \href{https://dx.doi.org/10.1088/2058-9565/ac1dff}{Quantum Sci.
  Technol. {\bf 6}, 045020}~(2021).
\newblock  \href{http://arxiv.org/abs/1910.09478}{arXiv:1910.09478}.

\bibitem{Yeter-Aydeniz:2020jte}
K\"ubra Yeter-Aydeniz, George Siopsis, and Raphael~C. Pooser.
\newblock ``{Scattering in the Ising model with the quantum Lanczos
  algorithm}''.
\newblock \href{https://dx.doi.org/10.1088/1367-2630/abe63d}{New J. Phys. {\bf
  23}, 043033}~(2021).
\newblock  \href{http://arxiv.org/abs/2008.08763}{arXiv:2008.08763}.

\bibitem{Milsted:2020jmf}
Ashley Milsted, Junyu Liu, John Preskill, and Guifre Vidal.
\newblock ``{Collisions of False-Vacuum Bubble Walls in a Quantum Spin
  Chain}''.
\newblock \href{https://dx.doi.org/10.1103/PRXQuantum.3.020316}{PRX Quantum
  {\bf 3}, 020316}~(2022).
\newblock  \href{http://arxiv.org/abs/2012.07243}{arXiv:2012.07243}.

\bibitem{Gustafson:2021imb}
Erik Gustafson, Yingyue Zhu, Patrick Dreher, Norbert~M. Linke, and Yannick
  Meurice.
\newblock ``{Real-time quantum calculations of phase shifts using wave packet
  time delays}''.
\newblock \href{https://dx.doi.org/10.1103/PhysRevD.104.054507}{Phys. Rev. D
  {\bf 104}, 054507}~(2021).
\newblock  \href{http://arxiv.org/abs/2103.06848}{arXiv:2103.06848}.

\bibitem{Deliyannis:2022uyh}
Plato Deliyannis, James Sud, Diana Chamaki, Zo\"e Webb-Mack, Christian~W.
  Bauer, and Benjamin Nachman.
\newblock ``{Improving quantum simulation efficiency of final state radiation
  with dynamic quantum circuits}''.
\newblock \href{https://dx.doi.org/10.1103/PhysRevD.106.036007}{Phys. Rev. D
  {\bf 106}, 036007}~(2022).
\newblock  \href{http://arxiv.org/abs/2203.10018}{arXiv:2203.10018}.

\bibitem{Dreher:2022scr}
Patrick Dreher, Erik Gustafson, Yingyue Zhu, Norbert~M. Linke, and Yannick
  Meurice.
\newblock ``{Real-time Quantum Calculations of Phase Shifts On NISQ Hardware
  Platforms Using Wavepacket Time Delay}''.
\newblock \href{https://dx.doi.org/10.22323/1.396.0464}{PoS {\bf LATTICE2021},
  464}~(2022).

\bibitem{Wang:2021iox}
Xiaoyang Wang, Xu~Feng, Lena Funcke, Tobias Hartung, Karl Jansen, Stefan
  K\"uhn, Georgios Polykratis, and Paolo Stornati.
\newblock ``{Model-Independent Error Mitigation in Parametric Quantum Circuits
  and Depolarizing Projection of Quantum Noise}''.
\newblock \href{https://dx.doi.org/10.22323/1.396.0603}{PoS {\bf LATTICE2021},
  603}~(2022).
\newblock  \href{http://arxiv.org/abs/2111.15522}{arXiv:2111.15522}.

\bibitem{Iannelli:2021jhs}
Giovanni Iannelli and Karl Jansen.
\newblock ``{Noisy Bayesian optimization for variational quantum
  eigensolvers}''.
\newblock \href{https://dx.doi.org/10.22323/1.396.0251}{PoS {\bf LATTICE2021},
  251}~(2022).
\newblock  \href{http://arxiv.org/abs/2112.00426}{arXiv:2112.00426}.

\bibitem{Yeter-Aydeniz:2022vuy}
K\"ubra Yeter-Aydeniz, Zachary Parks, Aadithya Nair, Erik Gustafson,
  Alexander~F. Kemper, Raphael~C. Pooser, Yannick Meurice, and Patrick Dreher.
\newblock ``{Measuring NISQ Gate-Based Qubit Stability Using a 1+1 Field Theory
  and Cycle Benchmarking}''~(2022).
\newblock  \href{http://arxiv.org/abs/2201.02899}{arXiv:2201.02899}.

\bibitem{Halimeh:2022mct}
Jad~C. Halimeh and Philipp Hauke.
\newblock ``{Stabilizing Gauge Theories in Quantum Simulators: A Brief
  Review}''.
\newblock In Proceedings of the 2021 Quantum Simulation for Strong Interactions
  (QuaSi) Workshops.
\newblock ~(2022).
\newblock  \href{http://arxiv.org/abs/2204.13709}{arXiv:2204.13709}.

\bibitem{Tuysuz:2022knj}
Cenk T\"uys\"uz, Giuseppe Clemente, Arianna Crippa, Tobias Hartung, Stefan
  K\"uhn, and Karl Jansen.
\newblock ``{Classical Splitting of Parametrized Quantum Circuits}''~(2022).
\newblock  \href{http://arxiv.org/abs/2206.09641}{arXiv:2206.09641}.

\bibitem{Jang:2022nun}
Wonho Jang, Koji Terashi, Masahiko Saito, Christian~W. Bauer, Benjamin Nachman,
  Yutaro Iiyama, Ryunosuke Okubo, and Ryu Sawada.
\newblock ``{Initial-State Dependent Optimization of Controlled Gate Operations
  with Quantum Computer}''~(2022).
\newblock  \href{http://arxiv.org/abs/2209.02322}{arXiv:2209.02322}.

\bibitem{Cubitt_2018}
Toby~S. Cubitt, Ashley Montanaro, and Stephen Piddock.
\newblock ``Universal quantum hamiltonians''.
\newblock \href{https://dx.doi.org/10.1073/pnas.1804949115}{Proceedings of the
  National Academy of Sciences {\bf 115}, 9497--9502}~(2018).

\bibitem{WILSON197475}
Kenneth~G. Wilson and J.~Kogut.
\newblock ``The renormalization group and the $\epsilon$ expansion''.
\newblock
  \href{https://dx.doi.org/https://doi.org/10.1016/0370-1573(74)90023-4}{Physics
  Reports {\bf 12}, 75--199}~(1974).

\bibitem{RevModPhys.55.583}
Kenneth~G. Wilson.
\newblock ``The renormalization group and critical phenomena''.
\newblock \href{https://dx.doi.org/10.1103/RevModPhys.55.583}{Rev. Mod. Phys.
  {\bf 55}, 583--600}~(1983).

\bibitem{RevModPhys.51.659}
John~B. Kogut.
\newblock ``An introduction to lattice gauge theory and spin systems''.
\newblock \href{https://dx.doi.org/10.1103/RevModPhys.51.659}{Rev. Mod. Phys.
  {\bf 51}, 659--713}~(1979).

\bibitem{CHANDRASEKHARAN2002388}
S.~Chandrasekharan, B.~Scarlet, and U.-J. Wiese.
\newblock ``From spin ladders to the 2d o(3) model at non-zero density''.
\newblock
  \href{https://dx.doi.org/https://doi.org/10.1016/S0010-4655(02)00311-9}{Computer
  Physics Communications {\bf 147}, 388--393}~(2002).

\bibitem{BROWER2004149}
R.~Brower, S.~Chandrasekharan, S.~Riederer, and U.-J. Wiese.
\newblock ``D-theory: field quantization by dimensional reduction of discrete
  variables''.
\newblock
  \href{https://dx.doi.org/https://doi.org/10.1016/j.nuclphysb.2004.06.007}{Nuclear
  Physics B {\bf 693}, 149--175}~(2004).

\bibitem{PhysRevD.99.074501}
Falk Bruckmann, Karl Jansen, and Stefan K\"uhn.
\newblock ``O(3) nonlinear sigma model in $1+1$ dimensions with matrix product
  states''.
\newblock \href{https://dx.doi.org/10.1103/PhysRevD.99.074501}{Phys. Rev. D
  {\bf 99}, 074501}~(2019).

\bibitem{PhysRevD.100.054505}
Hersh Singh and Shailesh Chandrasekharan.
\newblock ``Qubit regularization of the $o(3)$ sigma model''.
\newblock \href{https://dx.doi.org/10.1103/PhysRevD.100.054505}{Phys. Rev. D
  {\bf 100}, 054505}~(2019).

\bibitem{PhysRevLett.126.172001}
Tanmoy Bhattacharya, Alexander~J. Buser, Shailesh Chandrasekharan, Rajan Gupta,
  and Hersh Singh.
\newblock ``Qubit regularization of asymptotic freedom''.
\newblock \href{https://dx.doi.org/10.1103/PhysRevLett.126.172001}{Phys. Rev.
  Lett. {\bf 126}, 172001}~(2021).

\bibitem{https://doi.org/10.48550/arxiv.1911.12353}
Hersh Singh.
\newblock ``Qubit regularized $o(n)$ nonlinear sigma models''~(2019).

\bibitem{https://doi.org/10.48550/arxiv.2203.00059}
Hersh Singh.
\newblock ``Large-charge conformal dimensions at the $o(n)$ wilson-fisher fixed
  point''~(2022).

\bibitem{https://doi.org/10.48550/arxiv.2203.15766}
Stephan Caspar and Hersh Singh.
\newblock ``From asymptotic freedom to $θ$ vacua: Qubit embeddings of the o(3)
  nonlinear $σ$ model''~(2022).

\bibitem{Childs_2018}
Andrew~M. Childs, Dmitri Maslov, Yunseong Nam, Neil~J. Ross, and Yuan Su.
\newblock ``Toward the first quantum simulation with quantum speedup''.
\newblock \href{https://dx.doi.org/10.1073/pnas.1801723115}{Proceedings of the
  National Academy of Sciences {\bf 115}, 9456--9461}~(2018).

\bibitem{Childs_2021}
Andrew~M. Childs, Yuan Su, Minh~C. Tran, Nathan Wiebe, and Shuchen Zhu.
\newblock ``Theory of trotter error with commutator scaling''.
\newblock \href{https://dx.doi.org/10.1103/PhysRevX.11.011020}{Phys. Rev. X
  {\bf 11}, 011020}~(2021).

\bibitem{PRXQuantum.2.020328}
Tasio Gonzalez-Raya, Rodrigo Asensio-Perea, Ana Martin, Lucas~C. C\'eleri,
  Mikel Sanz, Pavel Lougovski, and Eugene~F. Dumitrescu.
\newblock ``Digital-analog quantum simulations using the cross-resonance
  effect''.
\newblock \href{https://dx.doi.org/10.1103/PRXQuantum.2.020328}{PRX Quantum
  {\bf 2}, 020328}~(2021).

\bibitem{PhysRevB.95.024431}
A.~Bermudez, L.~Tagliacozzo, G.~Sierra, and P.~Richerme.
\newblock ``Long-range heisenberg models in quasiperiodically driven crystals
  of trapped ions''.
\newblock \href{https://dx.doi.org/10.1103/PhysRevB.95.024431}{Phys. Rev. B
  {\bf 95}, 024431}~(2017).

\bibitem{PRXQuantum.3.020303}
P.~Scholl, H.~J. Williams, G.~Bornet, F.~Wallner, D.~Barredo, L.~Henriet,
  A.~Signoles, C.~Hainaut, T.~Franz, S.~Geier, A.~Tebben, A.~Salzinger,
  G.~Z\"urn, T.~Lahaye, M.~Weidem\"uller, and A.~Browaeys.
\newblock ``Microwave engineering of programmable $xxz$ hamiltonians in arrays
  of rydberg atoms''.
\newblock \href{https://dx.doi.org/10.1103/PRXQuantum.3.020303}{PRX Quantum
  {\bf 3}, 020303}~(2022).

\bibitem{MADI1997300}
Z.L. Mádi, B.~Brutscher, T.~Schulte-Herbrüggen, R.~Brüschweiler, and R.R.
  Ernst.
\newblock ``Time-resolved observation of spin waves in a linear chain of
  nuclear spins''.
\newblock
  \href{https://dx.doi.org/https://doi.org/10.1016/S0009-2614(97)00194-2}{Chemical
  Physics Letters {\bf 268}, 300--305}~(1997).

\bibitem{PhysRevLett.124.030601}
C.~M. S\'anchez, A.~K. Chattah, K.~X. Wei, L.~Buljubasich, P.~Cappellaro, and
  H.~M. Pastawski.
\newblock ``Perturbation independent decay of the loschmidt echo in a many-body
  system''.
\newblock \href{https://dx.doi.org/10.1103/PhysRevLett.124.030601}{Phys. Rev.
  Lett. {\bf 124}, 030601}~(2020).

\bibitem{https://doi.org/10.48550/arxiv.2207.09438}
Anthony~N. Ciavarella, Stephan Caspar, Hersh Singh, Martin~J. Savage, and Pavel
  Lougovski.
\newblock ``Floquet engineering heisenberg from ising using constant drive
  fields for quantum simulation''~(2022).

\bibitem{Richerme_2014}
Philip Richerme, Zhe-Xuan Gong, Aaron Lee, Crystal Senko, Jacob Smith, Michael
  Foss-Feig, Spyridon Michalakis, Alexey~V. Gorshkov, and Christopher Monroe.
\newblock ``Non-local propagation of correlations in quantum systems with
  long-range interactions''.
\newblock \href{https://dx.doi.org/10.1038/nature13450}{Nature {\bf 511},
  198--201}~(2014).

\bibitem{Jurcevic_2014}
P.~Jurcevic, B.~P. Lanyon, P.~Hauke, C.~Hempel, P.~Zoller, R.~Blatt, and C.~F.
  Roos.
\newblock ``Quasiparticle engineering and entanglement propagation in a quantum
  many-body system''.
\newblock \href{https://dx.doi.org/10.1038/nature13461}{Nature {\bf 511},
  202--205}~(2014).

\bibitem{Wall_2017}
Michael~L. Wall, Arghavan Safavi-Naini, and Ana~Maria Rey.
\newblock ``Boson-mediated quantum spin simulators in transverse fields: $xy$
  model and spin-boson entanglement''.
\newblock \href{https://dx.doi.org/10.1103/PhysRevA.95.013602}{Phys. Rev. A
  {\bf 95}, 013602}~(2017).

\bibitem{Kiely_2018}
Thomas~G. Kiely and J.~K. Freericks.
\newblock ``Relationship between the transverse-field ising model and the $xy$
  model via the rotating-wave approximation''.
\newblock \href{https://dx.doi.org/10.1103/PhysRevA.97.023611}{Phys. Rev. A
  {\bf 97}, 023611}~(2018).

\bibitem{https://doi.org/10.48550/arxiv.2208.01869}
Jeremy~T. Young, Sean~R. Muleady, Michael~A. Perlin, Adam~M. Kaufman, and
  Ana~Maria Rey.
\newblock ``Enhancing spin squeezing using soft-core interactions''~(2022).

\bibitem{abanin2017rigorous}
Dmitry Abanin, Wojciech De~Roeck, Wen~Wei Ho, and Fran{\c{c}}ois Huveneers.
\newblock ``A rigorous theory of many-body prethermalization for periodically
  driven and closed quantum systems''.
\newblock \href{https://dx.doi.org/10.1007/s00220-017-2930-x}{Commun. Math.
  Phys. {\bf 354}, 809--827}~(2017).

\bibitem{PhysRevX.7.011026}
Dominic~V. Else, Bela Bauer, and Chetan Nayak.
\newblock ``Prethermal phases of matter protected by time-translation
  symmetry''.
\newblock \href{https://dx.doi.org/10.1103/PhysRevX.7.011026}{Phys. Rev. X {\bf
  7}, 011026}~(2017).

\bibitem{PhysRevB.100.020301}
Kaoru Mizuta, Kazuaki Takasan, and Norio Kawakami.
\newblock ``High-frequency expansion for floquet prethermal phases with
  emergent symmetries: Application to time crystals and floquet engineering''.
\newblock \href{https://dx.doi.org/10.1103/PhysRevB.100.020301}{Phys. Rev. B
  {\bf 100}, 020301}~(2019).

\bibitem{https://doi.org/10.48550/arxiv.2103.07485}
Martin Claassen.
\newblock ``Flow renormalization and emergent prethermal regimes of
  periodically-driven quantum systems''~(2021).

\bibitem{https://doi.org/10.48550/arxiv.2203.01948}
Ieva Čepaitė, Anatoli Polkovnikov, Andrew~J. Daley, and Callum~W. Duncan.
\newblock ``Counterdiabatic optimised local driving''~(2022).

\bibitem{saffman2010quantum}
Mark Saffman, Thad~G Walker, and Klaus M{\o}lmer.
\newblock ``Quantum information with rydberg atoms''.
\newblock \href{https://dx.doi.org/10.1103/RevModPhys.82.2313}{Rev. Mod. Phys.
  {\bf 82}, 2313}~(2010).

\bibitem{bernien2017probing}
Hannes Bernien, Sylvain Schwartz, Alexander Keesling, Harry Levine, Ahmed
  Omran, Hannes Pichler, Soonwon Choi, Alexander~S Zibrov, Manuel Endres,
  Markus Greiner, et~al.
\newblock ``Probing many-body dynamics on a 51-atom quantum simulator''.
\newblock \href{https://dx.doi.org/10.1038/nature24622}{Nature {\bf 551},
  579--584}~(2017).

\bibitem{PhysRevX.8.021070}
Vincent Lienhard, Sylvain de~L\'es\'eleuc, Daniel Barredo, Thierry Lahaye,
  Antoine Browaeys, Michael Schuler, Louis-Paul Henry, and Andreas~M.
  L\"auchli.
\newblock ``Observing the space- and time-dependent growth of correlations in
  dynamically tuned synthetic ising models with antiferromagnetic
  interactions''.
\newblock \href{https://dx.doi.org/10.1103/PhysRevX.8.021070}{Phys. Rev. X {\bf
  8}, 021070}~(2018).

\bibitem{PhysRevX.8.021069}
Elmer Guardado-Sanchez, Peter~T. Brown, Debayan Mitra, Trithep Devakul,
  David~A. Huse, Peter Schau\ss{}, and Waseem~S. Bakr.
\newblock ``Probing the quench dynamics of antiferromagnetic correlations in a
  2d quantum ising spin system''.
\newblock \href{https://dx.doi.org/10.1103/PhysRevX.8.021069}{Phys. Rev. X {\bf
  8}, 021069}~(2018).

\bibitem{kim2018detailed}
Hyosub Kim, YeJe Park, Kyungtae Kim, H-S Sim, and Jaewook Ahn.
\newblock ``Detailed balance of thermalization dynamics in rydberg-atom quantum
  simulators''.
\newblock \href{https://dx.doi.org/10.1103/PhysRevLett.120.180502}{Phy. Rev.
  Lett. {\bf 120}, 180502}~(2018).

\bibitem{de2019observation}
Sylvain de~L{\'e}s{\'e}leuc, Vincent Lienhard, Pascal Scholl, Daniel Barredo,
  Sebastian Weber, Nicolai Lang, Hans~Peter B{\"u}chler, Thierry Lahaye, and
  Antoine Browaeys.
\newblock ``Observation of a symmetry-protected topological phase of
  interacting bosons with rydberg atoms''.
\newblock \href{https://dx.doi.org/10.1126/science.aav9105}{Science {\bf 365},
  775--780}~(2019).

\bibitem{keesling2019quantum}
Alexander Keesling, Ahmed Omran, Harry Levine, Hannes Bernien, Hannes Pichler,
  Soonwon Choi, Rhine Samajdar, Sylvain Schwartz, Pietro Silvi, Subir Sachdev,
  et~al.
\newblock ``Quantum kibble--zurek mechanism and critical dynamics on a
  programmable rydberg simulator''.
\newblock \href{https://dx.doi.org/10.1038/s41586-019-1070-1}{Nature {\bf 568},
  207--211}~(2019).

\bibitem{morgado2021quantum}
M~Morgado and S~Whitlock.
\newblock ``Quantum simulation and computing with rydberg-interacting qubits''.
\newblock \href{https://dx.doi.org/10.1116/5.0036562}{AVS Quantum Sci. {\bf 3},
  023501}~(2021).

\bibitem{semeghini2021probing}
Giulia Semeghini, Harry Levine, Alexander Keesling, Sepehr Ebadi, Tout~T Wang,
  Dolev Bluvstein, Ruben Verresen, Hannes Pichler, Marcin Kalinowski, Rhine
  Samajdar, et~al.
\newblock ``Probing topological spin liquids on a programmable quantum
  simulator''.
\newblock \href{https://dx.doi.org/10.1126/science.abi8794}{Science {\bf 374},
  1242--1247}~(2021).

\bibitem{ebadi2021quantum}
Sepehr Ebadi, Tout~T Wang, Harry Levine, Alexander Keesling, Giulia Semeghini,
  Ahmed Omran, Dolev Bluvstein, Rhine Samajdar, Hannes Pichler, Wen~Wei Ho,
  et~al.
\newblock ``Quantum phases of matter on a 256-atom programmable quantum
  simulator''.
\newblock \href{https://dx.doi.org/10.1038/s41586-021-03582-4}{Nature {\bf
  595}, 227--232}~(2021).

\bibitem{masanes2002time}
Ll~Masanes, Guifr{\'e} Vidal, and Jos{\'e}~Ignacio Latorre.
\newblock ``Time-optimal hamiltonian simulation and gate synthesis using
  homogeneous local unitaries''~(2002).

\bibitem{gardas2018defects}
Bart{\l}omiej Gardas, Jacek Dziarmaga, Wojciech~H Zurek, and Michael Zwolak.
\newblock ``Defects in quantum computers''.
\newblock \href{https://dx.doi.org/10.1038/s41598-018-22763-2}{Sci. Rep. {\bf
  8}, 1--10}~(2018).

\bibitem{doi:10.1126/science.aat2025}
R.~Harris, Y.~Sato, A.~J. Berkley, M.~Reis, F.~Altomare, M.~H. Amin,
  K.~Boothby, P.~Bunyk, C.~Deng, C.~Enderud, S.~Huang, E.~Hoskinson, M.~W.
  Johnson, E.~Ladizinsky, N.~Ladizinsky, T.~Lanting, R.~Li, T.~Medina,
  R.~Molavi, R.~Neufeld, T.~Oh, I.~Pavlov, I.~Perminov, G.~Poulin-Lamarre,
  C.~Rich, A.~Smirnov, L.~Swenson, N.~Tsai, M.~Volkmann, J.~Whittaker, and
  J.~Yao.
\newblock ``Phase transitions in a programmable quantum spin glass simulator''.
\newblock \href{https://dx.doi.org/10.1126/science.aat2025}{Science {\bf 361},
  162--165}~(2018).

\bibitem{King_2018}
Andrew~D. King, Juan Carrasquilla, Jack Raymond, Isil Ozfidan, Evgeny
  Andriyash, Andrew Berkley, Mauricio Reis, Trevor Lanting, Richard Harris,
  Fabio Altomare, Kelly Boothby, Paul~I. Bunyk, Colin Enderud, Alexandre
  Fr{\'{e}}chette, Emile Hoskinson, Nicolas Ladizinsky, Travis Oh, Gabriel
  Poulin-Lamarre, Christopher Rich, Yuki Sato, Anatoly~Yu. Smirnov, Loren~J.
  Swenson, Mark~H. Volkmann, Jed Whittaker, Jason Yao, Eric Ladizinsky, Mark~W.
  Johnson, Jeremy Hilton, and Mohammad~H. Amin.
\newblock ``Observation of topological phenomena in a programmable lattice of
  1,800 qubits''.
\newblock \href{https://dx.doi.org/10.1038/s41586-018-0410-x}{Nature {\bf 560},
  456--460}~(2018).

\bibitem{PhysRevLett.124.090502}
Phillip Weinberg, Marek Tylutki, Jami~M. R\"onkk\"o, Jan Westerholm, Jan~A.
  \AA{}str\"om, Pekka Manninen, P\"aivi T\"orm\"a, and Anders~W. Sandvik.
\newblock ``Scaling and diabatic effects in quantum annealing with a d-wave
  device''.
\newblock \href{https://dx.doi.org/10.1103/PhysRevLett.124.090502}{Phys. Rev.
  Lett. {\bf 124}, 090502}~(2020).

\bibitem{King_2021}
Andrew~D. King, Jack Raymond, Trevor Lanting, Sergei~V. Isakov, Masoud Mohseni,
  Gabriel Poulin-Lamarre, Sara Ejtemaee, William Bernoudy, Isil Ozfidan,
  Anatoly~Yu. Smirnov, Mauricio Reis, Fabio Altomare, Michael Babcock, Catia
  Baron, Andrew~J. Berkley, Kelly Boothby, Paul~I. Bunyk, Holly Christiani,
  Colin Enderud, Bram Evert, Richard Harris, Emile Hoskinson, Shuiyuan Huang,
  Kais Jooya, Ali Khodabandelou, Nicolas Ladizinsky, Ryan Li, P.~Aaron Lott,
  Allison J.~R. MacDonald, Danica Marsden, Gaelen Marsden, Teresa Medina, Reza
  Molavi, Richard Neufeld, Mana Norouzpour, Travis Oh, Igor Pavlov, Ilya
  Perminov, Thomas Prescott, Chris Rich, Yuki Sato, Benjamin Sheldan, George
  Sterling, Loren~J. Swenson, Nicholas Tsai, Mark~H. Volkmann, Jed~D.
  Whittaker, Warren Wilkinson, Jason Yao, Hartmut Neven, Jeremy~P. Hilton, Eric
  Ladizinsky, Mark~W. Johnson, and Mohammad~H. Amin.
\newblock ``Scaling advantage over path-integral monte carlo in quantum
  simulation of geometrically frustrated magnets''.
\newblock \href{https://dx.doi.org/10.1038/s41467-021-20901-5}{Nat. Commun.{\bf
  12}}~(2021).

\bibitem{PhysRevResearch.2.033369}
Yuki Bando, Yuki Susa, Hiroki Oshiyama, Naokazu Shibata, Masayuki Ohzeki,
  Fernando~Javier G\'omez-Ruiz, Daniel~A. Lidar, Sei Suzuki, Adolfo del Campo,
  and Hidetoshi Nishimori.
\newblock ``Probing the universality of topological defect formation in a
  quantum annealer: Kibble-zurek mechanism and beyond''.
\newblock \href{https://dx.doi.org/10.1103/PhysRevResearch.2.033369}{Phys. Rev.
  Research {\bf 2}, 033369}~(2020).

\bibitem{PRXQuantum.1.020320}
Paul Kairys, Andrew~D. King, Isil Ozfidan, Kelly Boothby, Jack Raymond, Arnab
  Banerjee, and Travis~S. Humble.
\newblock ``Simulating the shastry-sutherland ising model using quantum
  annealing''.
\newblock \href{https://dx.doi.org/10.1103/PRXQuantum.1.020320}{PRX Quantum
  {\bf 1}, 020320}~(2020).

\bibitem{https://doi.org/10.48550/arxiv.2003.14244}
T.~Lanting, M.~H. Amin, C.~Baron, M.~Babcock, J.~Boschee, S.~Boixo, V.~N.
  Smelyanskiy, M.~Foygel, and A.~G. Petukhov.
\newblock ``Probing environmental spin polarization with superconducting flux
  qubits''~(2020).

\bibitem{PhysRevA.102.042403}
Kohji Nishimura, Hidetoshi Nishimori, and Helmut~G. Katzgraber.
\newblock ``Griffiths-mccoy singularity on the diluted chimera graph: Monte
  carlo simulations and experiments on quantum hardware''.
\newblock \href{https://dx.doi.org/10.1103/PhysRevA.102.042403}{Phys. Rev. A
  {\bf 102}, 042403}~(2020).

\bibitem{doi:10.1126/science.abe2824}
Andrew~D. King, Cristiano Nisoli, Edward~D. Dahl, Gabriel Poulin-Lamarre, and
  Alejandro Lopez-Bezanilla.
\newblock ``Qubit spin ice''.
\newblock \href{https://dx.doi.org/10.1126/science.abe2824}{Science {\bf 373},
  576--580}~(2021).

\bibitem{https://doi.org/10.48550/arxiv.2202.05847}
Andrew~D. King, Sei Suzuki, Jack Raymond, Alex Zucca, Trevor Lanting, Fabio
  Altomare, Andrew~J. Berkley, Sara Ejtemaee, Emile Hoskinson, Shuiyuan Huang,
  Eric Ladizinsky, Allison MacDonald, Gaelen Marsden, Travis Oh, Gabriel
  Poulin-Lamarre, Mauricio Reis, Chris Rich, Yuki Sato, Jed~D. Whittaker, Jason
  Yao, Richard Harris, Daniel~A. Lidar, Hidetoshi Nishimori, and Mohammad~H.
  Amin.
\newblock ``{Coherent quantum annealing in a programmable 2,000\,qubit Ising
  chain}''.
\newblock \href{https://dx.doi.org/10.1038/s41567-022-01741-6}{Nature Phys.
  {\bf 18}, 1324--1328}~(2022).

\end{thebibliography}

%%%%%%%%%%%%%%%%%%%%%%%%
\clearpage
\appendix

%%%%%%%%%%%%%%%%%%%%%%%%%

\end{document}